\documentclass[journal]{IEEEtran}

\ifCLASSOPTIONcompsoc
  \usepackage[nocompress]{cite}
\else
  \usepackage{cite}
\fi

\ifCLASSINFOpdf
\else
\fi

\usepackage[loading]{tracefnt}
\usepackage{algorithmicx}
\usepackage{algorithm}
\usepackage{algpseudocode}
\usepackage{multirow}
\usepackage{subcaption}
\usepackage{setspace}
\usepackage[utf8]{inputenc}
\usepackage{amsmath}
\usepackage{amsfonts}
\usepackage{amssymb}
\usepackage{graphicx}
\usepackage{color}
\usepackage{hyperref}

\renewcommand{\figurename}{Figure}
\newcommand{\RN}[1]{%
  \textup{\uppercase\expandafter{\romannumeral#1}}%
}
\begin{document}
%
\title{Formal Analysis of Galois Field Arithmetic Circuits\\ {\LARGE -$~$Parallel Verification and Reverse Engineering}}

\author{Cunxi~Yu~\IEEEmembership{Student Member,~IEEE,}
        and Maciej~Ciesielski,~\IEEEmembership{Senior Member,~IEEE}
\IEEEcompsocitemizethanks{\IEEEcompsocthanksitem C. Yu and M. Ciesielski are with the Department
of Electrical and Computer Engineering, University of Massachusetts, Amherst, MA, 01375. The related tools and benchmarks are released publicly on Github, \textit{ycunxi.github.io/Parallel\_Formal\_Analysis\_GaloisField\/}\protect\\
E-mail: ycunxi@umass.edu}}

\markboth{IEEE TRANSACTIONS ON COMPUTER-AIDED DESIGN OF INTEGRATED CIRCUITS AND SYSTEMS}%
{Shell \MakeLowercase{\textit{et al.}}: Bare Demo of IEEEtran.cls for Computer Society Journals}

\IEEEtitleabstractindextext{%
\begin{abstract}
Galois field (GF) arithmetic circuits find numerous applications in communications, signal processing, and security engineering. Formal verification techniques of GF circuits are scarce and limited to circuits with known bit positions of the primary inputs and outputs. They also require knowledge of the irreducible polynomial $P(x)$, which affects final hardware implementation. This paper presents a computer algebra technique that performs verification and reverse engineering of GF($2^m$) multipliers directly from the gate-level implementation. The approach is based on extracting a unique irreducible polynomial in a parallel fashion and proceeds in three steps: 1) determine the bit position of the output bits; 2) determine the bit position of the input bits; and 3) extract the irreducible polynomial used in the design. We demonstrate that this method is able to reverse engineer GF($2^m$) multipliers in \textit{m} threads. Experiments performed on synthesized \textit{Mastrovito} and \textit{Montgomery} multipliers with different $P(x)$, including NIST-recommended polynomials, demonstrate high efficiency of the proposed method.
\end{abstract}

}

\maketitle

\IEEEdisplaynontitleabstractindextext

%
\IEEEpeerreviewmaketitle

\section{Introduction} \label{sec:introduction}

\IEEEPARstart{D}{espite} considerable progress in verification of random and control logic, advances in formal verification of arithmetic circuits have been lagging. This can be attributed to the difficulty in efficient modeling of arithmetic circuits and datapaths, without resorting to computationally expensive Boolean methods. 
Contemporary formal techniques, such as \textit{Binary Decision Diagrams} (BDDs), \textit{Boolean Satisfiability} (SAT), \textit{Satisfiability Modulo Theories} (SMT), etc., are not directly applicable to verification  of integer and finite field arithmetic circuits \cite{kalla:tcad13}\cite{ciesielski2015verification}. 
This paper concentrates on formal verification and reverse engineering of finite (Galois) field arithmetic circuits.

Galois field (GF) is a number system with a finite number of elements and two main arithmetic operations, addition and multiplication; other operations can be derived from those two \cite{paar2009understanding}. 
{ 
GF arithmetic plays an important role in coding theory, cryptography, and their numerous applications. Therefore, developing formal techniques for hardware implementations of GF arithmetic circuits, and particularly for finite field multiplication, is essential.}

The elements in field GF($2^m$) can be represented using polynomial rings. The field of size $m$ is constructed using \textit{irreducible polynomial} $P(x)$, which includes terms of degree with $d$ $\in$ [$0, m$] with coefficients in GF(2). The arithmetic operation in the field is then performed modulo $P(x)$. 
The choice of the irreducible polynomial has a significant impact on the hardware implementation of the GF circuit and its performance. Typically, the irreducible polynomial with a minimum number of elements gives the best performance \cite{ciet2002short}, but it is not always the case.

Due to the rising number of threats in hardware security, analyzing finite field circuits becomes important. Computer algebra techniques with polynomial representation seem to offer the best solution for analyzing arithmetic circuits. Several works address the verification and functional abstraction problems, both in Galois field arithmetic  \cite{kalla:tcad13}\cite{kalla:dac2014}\cite{PrussKE16TCAD} and integer arithmetic implementations \cite{STABLE:date11}\cite{ciesielski2015verification}\cite{farahmandi2015groebner}\cite{sayedformal:date-2016}\cite{yu:2016-tcad-verification}. Symbolic computer algebra methods have also been used to reverse engineer the word-level operations for GF circuits and integer arithmetic circuits to improve verification performance \cite{yu:2016-abstraction}\cite{sayedequivalence}\cite{kalla:dac2014}. The verification problem is typically formulated as proving that the implementation satisfies the specification, and is accomplished by performing a series of divisions of the specification polynomial by the implementation polynomials. In the work of Yu {\it et al.} \cite{yu:2016-abstraction}, the authors proposed an original spectral method based on analyzing the internal algebraic expressions during the rewriting procedure. Sayed-Ahmed et al. \cite{sayedequivalence} introduced a reverse engineering technique in Algebraic Combinational Equivalence Checking (ACEC) process by converting the function into canonical polynomials and using \textit{Gr{\" o}bner Basis}. 

However, the above mentioned algebraic techniques have several limitations. Firstly, they are restricted to implementations with a known binary encoding of the inputs and outputs. This information is needed to generate the specification polynomial that describes the circuit functionality regarding its inputs and outputs, necessary for the polynomial reduction process (described in Section \ref{sec:computer-algebra}).
Secondly, these methods are unable to explore parallelism (inherent in GF circuits), 
as they require that the polynomial division is applied iteratively using reverse-topological order  \cite{ciesielski2015verification}\cite{sayedformal:date-2016}\cite{PrussKE16TCAD}. 
Thirdly, the approaches applied specifically to GF($2^m$) arithmetic circuits \cite{kalla:dac2014}\cite{PrussKE16TCAD}, require knowledge of the irreducible polynomial $P(x)$ of the circuit. 

In this work, we present a formal approach to reverse engineer the gate-level finite field arithmetic circuits that exploit inherent parallelism of the GF circuits. The method is based on a parallel algebraic rewriting approach \cite{yu:aspdac17} and applied specifically to multipliers. The objective of reverse engineering is as follows: given the netlist of a gate-level GF multiplier, extract the bit positions of input and output bits and the irreducible polynomial used in constructing the GF multiplication; then extract the specification of the design using this information. Bit position $i$ indicates the location of the bit in the binary word according to its significance (LSB vs MSB).
Our approach solves this problem by transforming the algebraic expressions of the output bits into an algebraic expression of the input bits (specification), and is done in parallel for each output bit. Specifically, it includes the following steps\footnote{Our tool and benchmarks used in this journal paper are released publicly at our project website at\\ \url{https://ycunxi.github.io/Parallel_Formal_Analysis_GaloisField}}:

\begin{itemize}
\item Extract the algebraic expression of each output bit.
\item Determine the bit position of the outputs.
\item Determine the bit position of the inputs.
\item Extract the irreducible polynomial $P(x)$.
\item Extract the specification by algebraic rewriting.
\end{itemize}

We demonstrate the efficiency of our method using GF($2^m$) \textit{Mastrovito} and \textit{Montgomery} multipliers of up to 571-bit width in a bit-blasted format (i.e., flattened to bit-level), implemented using various irreducible polynomials.


\section{Background} \label{sec:background}

\subsection{Canonical Diagrams}
Several approaches have been proposed to check an arithmetic circuit against its functional specification. Different variants of canonical, graph-based representations have been proposed, including Binary Decision Diagrams (BDDs) \cite{bryant:1986-bdd}, Binary Moment Diagrams (BMDs) \cite{bmd95} \cite{bryant:tr97}, Taylor Expansion Diagrams (TED) \cite{ted:tcomp06}, and other hybrid diagrams. 
While BDDs have been used extensively in logic synthesis, their application to verification of arithmetic circuits is limited by the prohibitively high memory requirement for complex arithmetic circuits, such as multipliers. 
BDDs are being used, along with many other methods, for local reasoning, but not as monolithic data structure \cite{kaivola:2009-intel}. BMDs and TEDs offer a better space complexity but require word-level information of the design, which is often not available or is hard to extract from bit-level netlists. While the canonical diagrams have been used extensively in logic synthesis, high-level synthesis, and verification, their application to verify large arithmetic circuits remains limited by the prohibitively high memory requirement of complex arithmetic circuits \cite{ciesielski2015verification}\cite{kalla:tcad13}.

\subsection{SAT, ILP and SMT Solvers}
Arithmetic verification problems have been typically modeled using Boolean satisfiability (SAT). Several SAT solvers have been developed to solve Boolean decision problems, including ABC \cite{mishchenko:abc-2007}, MiniSAT \cite{sorensson:2005-minisat}, and others. Some of them, such as CryptoMiniSAT \cite{soos:PoS-2010}, target specifically {\sc xor}-rich circuits, and are potentially useful for arithmetic circuit verification, but are all based on a computationally expensive DPLL (Davis, Putnam, Logemann, Loveland) decision procedure \cite{davis1962machine}. 
Some techniques combine automatic test pattern generation (ATPG) and modular arithmetic constraint-solving techniques for the purpose of test generation and assertion checking \cite{Cheng:tcad01}. 
Others integrate linear arithmetic constraints with Boolean SAT in a unified algebraic domain \cite{hsat:dac98}, but their effectiveness is limited by constraint propagation across the Boolean and word-level boundary. To avoid this problem, methods based on ILP models of arithmetic operators have been proposed \cite{Brinkman:aspdac02} \cite{LPSAT:jsa05}, but in general ILP techniques are known to be computationally expensive and not scalable to large scale systems.
\textit{SMT solvers} depart from treating the problem in a strictly Boolean domain and integrate different well-defined theories (Boolean logic, bit vectors, integer arithmetic, etc.) into a DPLL-style SAT decision procedure \cite{SMT-book:2008}. 
Some of the most effective SMT solvers, potentially applicable to our problem, are Boolector \cite{niemetz:2015boolector}, Z3 \cite{de:2008-z3}, and CVC \cite{barrett2011cvc4}. However, SMT solvers still model functional verification as a decision problem and, 
as demonstrated by extensive experimental results, neither SAT nor SMT solvers can efficiently solve the verification problem of large arithmetic circuits \cite{kalla:tcad13} \cite{yu:2016-tcad-verification}.


\subsection{Theorem Provers}
Another class of solvers include Theorem Provers, deductive systems for proving that an implementation satisfies the specification, using mathematical reasoning. The proof system is based on a large and strongly problem-specific database of axioms and inference rules, such as simplification, term rewriting, induction, etc. Some of the most popular theorem proving systems are: HOL \cite{gordon:1993-introduction}, PVS \cite{owre1992pvs}, ACL2 \cite{brock:1996-acl2}, and the term rewriting method described in \cite{Vasudevan:tcomp07}.
These systems are characterized by high abstraction and powerful logic expressiveness, but they are highly interactive, require intimate domain knowledge, extensive user guidance, and expertise for efficient use. 
The success of verification using theorem provers depends on the set of available axioms and rewrite rules, and on the choice and order in which the rules are applied during the proof process, with no guarantee for a conclusive answer \cite{kapur:1998-rrl}. 

\subsection{Computer Algebra Approaches} \label{sec:computer-algebra}

The most advanced techniques that have potential to solve the arithmetic verification problems are those based on Symbolic Computer Algebra. 
The verification problem is typically formulated as a proof that the implementation satisfies the specification \cite{ciesielski2015verification}\cite{kalla:tcad13}\cite{farahmandi2015groebner}\cite{STABLE:date11}\cite{sayedformal:date-2016}. This is accomplished by performing a series of divisions of the specification polynomial $F$ by a set of implementation polynomials $B$, representing circuit components, the process referred to as reduction of $F$ modulo $B$. 
Polynomials $f_1,...,f_s \in B$ are called the bases, or {\it generators}, of the ideal $J$. 
Given a set $f_1,...,f_s$ of generators of $J$, a set of all simultaneous solutions to a system of equations $f_1$=0; ... ,$f_s$=0 is called a {\it variety} $V(J)$. Verification problem is then formulated as a test if the specification $F$ vanishes on $V(J)$, i.e., if  $F \in V(J)$. 
This is known in computer algebra as {\it ideal membership} testing problem \cite{kalla:tcad13}.

Standard procedure to test if $F \in V(J)$ is to divide polynomial $F$ by the polynomials \{$f_1,...,f_s$\} of $B$, one by one. The goal is to cancel, at each iteration, the leading term of $F$ using one of the leading terms of $f_1,...,f_s$. If the remainder $r$ of the division is 0, then $F$ vanishes on $V(J)$, proving that the implementation satisfies the specification. 
However, if $ r \ne 0 $, such a conclusion cannot be made;
$B$ may not be sufficient to reduce $F$ to 0, and yet the circuit may be correct. To reliably check if $F$ is reducible to zero, a {\it canonical} set of generators, $G=\{g_1,...,g_t\}$, called {\it Gr{\" o}bner basis}, is needed. 
It has been shown that for combinational circuits with no feedback, certain conditions automatically make the set $B$ a Groebner basis \cite{stoffel:tcad04}. Specifically, if the polynomials $f_1,...,f_s \in B$ are ordered in reverse topological order of logic gates, from primary outputs to primary inputs, and the leading term of each polynomial is the output of a logic gate, then set $B$ is automatically a Groebner basis.
Some of the authors use Gaussian elimination, rather than explicit polynomial division, to speed up the polynomial reduction process \cite{kalla:tcad13}\cite{farahmandi2015groebner}. 
The polynomials corresponding to fanout-free logic cones can be precomputed to reduce the size of the problem \cite{farahmandi2015groebner}. 

The polynomial reduction technique has been successfully applied to both  integer arithmetic circuits \cite{sayedformal:date-2016} and Galois field arithmetic \cite{kalla:tcad13}.
Verification work of Galois field arithmetic has been presented in \cite{kalla:tcad13} \cite{kalla:dac2014}. Formulation of  problems in GF arithmetic takes advantage of known properties of Galois field during polynomial reductions. Specifically, the problem reduces to the ideal membership testing over a larger ideal that includes ideal $J_0 = \langle x^2-x \rangle $ in ${\mathbb{F}}_2$, for each internal signal $x$ of the circuit. Inclusion of this ideal basically assures that each signal assumes a binary value.
In this paper, we provide comparison between this technique and our approach.

\subsection{Function Extraction}

\textit{Function extraction} is an arithmetic verification method originally proposed in \cite{ciesielski2015verification} for arithmetic circuits in modular integer arithmetic $\mathbb{Z}_{2^m}$. 
It extracts a unique bit-level polynomial function implemented by the circuit directly from its gate-level implementation. 
Instead of expensive polynomial division, extraction is done by \textit{backward rewriting}, i.e., transforming the polynomial representing encoding of the primary outputs (called the \textit{output signature}) into a polynomial at the primary inputs (the \textit{input signature}) using algebraic models of the logic gates of the circuit. That is, the rewriting is performed in a reverse topological order.
This technique has been successfully applied to large integer arithmetic circuits, such as 512-bit integer multipliers.
However, it is not directly applicable to large Galois Field multipliers because of potentially exponential number of polynomial terms, before the internal term cancellations takes place during rewriting. 
Fortunately, arithmetic GF($2^{m}$) circuits offer an inherent parallelism which can be exploited in backward rewriting, without memory explosion.

In the rest of the paper, we first describe how to apply such parallel rewriting in GF($2^{m}$) circuits while avoiding memory explosion experienced in integer arithmetic circuits. Using this approach, we extract the function of each output bit in $\mathbb{F}_{2^m}$ and the function is represented in a {\it pseudo-Boolean polynomial} expression, where all variables are Boolean. Finally, we propose a method to reverse engineer the GF($2^m$) designs by analyzing these expressions.

\section{Galois Field Multiplication} \label{GF-multiplication}

Galois field (GF) is a number system with a finite number of elements and two main arithmetic operations, addition and multiplication; other operations such as division can be derived from those two \cite{paar2009understanding}. Galois field with $p$ elements is denoted as GF($p$). The most widely-used finite fields are \textit{Prime Fields} and \textit{Extension Fields}, and particularly {\it Binary Extension Fields}. Prime field, denoted GF($p$), is a finite field consisting of finite number of integers \{$1,2, ....,p-1$\}, where $p$ is a prime number, with additions and multiplication performed \textit{modulo p}. 
Binary extension field, denoted GF($2^m$) (or $\mathbb{F}_{2^m}$), is a finite field with $2^m$ elements. Unlike in prime fields, however, the operations in extension fields are not computed \textit{modulo $2^{m}$}. Instead, in one possible representation (called \textit{polynomial basis}), 
each element  of GF($2^m$) is a {\it polynomial ring} with $m$ terms with coefficients in GF(2), modulo $P(x)$.  Addition of field elements is the usual addition of polynomials, with coefficient arithmetic performed modulo 2.  
Multiplication of field elements is performed modulo {\it irreducible polynomial} $P(x)$ of degree $m$ and coefficients in GF(2). The irreducible polynomial $P(x)$ is analogous to the prime number $p$ in prime fields $GF(p)$. 
{ In this work, we focus on the verification problem of GF($2^m$) multipliers that appear in many cryptography and in some DSP applications.} 


\subsection{GF Multiplication Principle}

Two different GF multiplication structures, constructed using different irreducible polynomials $P_{1}(x)$ and $P_{2}(x)$, are shown in Figure \ref{fig:4-bit-gf-multiplication}. The integer multiplication takes two $n$-bit operands as input and generates a $2n$-bit word, where the values computed at lower significant bits ripple through the carry chain all the way to the most significant bit (MSB). In contrast, in GF($2^m$) implementations the number of outputs is reduced to $n$ using irreducible polynomial P(x). The product terms are added for each column (output bit position) modulo 2, hence there is no carry propagation. For example, to represent the result in GF{($2^{4}$)}, with only four output bits, the four most significant bits in the result of the integer multiplication have to be reduced to GF($2^4$). The result of such a reduction is shown in Figure \ref{fig:4-bit-gf-multiplication}. In GF($2^4$), the input and output operands are represented using polynomials $A(x)$, $B(x)$ and $Z(x)$, where $A(x)$=$\sum_{n=0}^{n=3} a_{n}\cdot x^{n} $, $B(x)$=$\sum_{n=0}^{n=3} b_{n}\cdot x^{n}$, and $Z(x)$=$\sum_{n=0}^{n=3} z_{n}\cdot x^{n}$, respectively. 

\textbf{Example 1:} The function of each multiplication bit $s_{i}$ ($i$ $\in$ [0, 6]) is represented using polynomials in GF(2), namely: $s_{0}$=$a_{0}b_{0}$, $s_{1}$=$a_{1}b_{0}$+$a_{0}b_{1}$, etc. ..., up to $s_{6}$=$a_{3}b_{3}$\footnote{For polynomials in GF(2), "+" are computed as modulo 2.}. The output bits $z_{n}$ ($n \in$ [0, 3]) are computed modulo the irreducible polynomial $P(x)$. Using $P_{2}(x)$=$x^4$+$x$+1, we obtain : $z_{0}$=$s_{0}$+$s_{4}$, $z_{1}$=$s_{1}$+$s_{4}$+$s_{5}$, $z_{2}$=$a_0$$b_2$+$a_1$$b_1$+$a_2$$b_0$+$a_2$$b_3$+$a_3$$b_2$+$a_3b_3$, and $z_{3}$=$a_0$$b_3$+$a_1$$b_2$+$a_2$$b_1$+$a_3$$b_0$+$a_3b_3$. 
The coefficients of the multiplication results are shown in Figure \ref{fig:4-bit}.  In digital circuits, partial products are implemented using {\sc and} gates, and addition modulo 2 is done using {\sc xor} gates. Note that, unlike in integer multiplication, in GF($2^m$) circuits there is no carry out to the next bit. For this reason, as we can see in Figure \ref{fig:4-bit-gf-multiplication}, the function of each output bit can be computed independently of other bits.

\begin{figure}[ht]
\centering
\begin{tabular}{lllllll}
             &              &              & $a_3$        & $a_2$        & $a_1$        & $a_0$        \\
             &              &              & $b_3$        & $b_2$        & $b_1$        & $b_0$        \\ \hline
             &              &              & $a_{3}b_{0}$ & $a_{2}b_{0}$ & $a_{1}b_{0}$ & $a_{0}b_{0}$ \\
             &              & $a_{3}b_{1}$ & $a_{2}b_{1}$ & $a_{1}b_{1}$ & $a_{0}b_{1}$ &              \\
             & $a_{3}b_{2}$ & $a_{2}b_{2}$ & $a_{1}b_{2}$ & $a_{0}b_{2}$ &              &              \\
$a_{3}b_{3}$ & $a_{2}b_{3}$ & $a_{1}b_{3}$ & $a_{0}b_{3}$ &              &              &              \\ \hline
$s_6$        & $s_5$        & $s_4$        & $s_3$        & $s_2$        & $s_1$        & $s_0$       
\end{tabular}
\vspace*{3mm}
\label{my-label}

        \begin{minipage}{.5\linewidth}
      \centering
\begin{tabular}{cccc}
\multicolumn{4}{l}{{$P_{1}(x)$=$x^{4}+x^{3}+1$}}                                                                            \\
\multicolumn{1}{c|}{$s_3$} & \multicolumn{1}{c|}{$s_2$} & \multicolumn{1}{c|}{$s_1$} & $s_0$                     \\
\multicolumn{1}{c|}{$s_4$} & \multicolumn{1}{c|}{0}  & \multicolumn{1}{c|}{0} & $s_4$                     \\
\multicolumn{1}{c|}{$s_5$} & \multicolumn{1}{c|}{0}  & \multicolumn{1}{c|}{$s_5$}  & $s_5$                     \\
\multicolumn{1}{c|}{$s_6$} & \multicolumn{1}{c|}{$s_6$} & \multicolumn{1}{c|}{$s_6$}  & $s_6$                      \\ \hline
\multicolumn{1}{l|}{$z_3$} & \multicolumn{1}{l|}{$z_2$} & \multicolumn{1}{l|}{$z_1$} & \multicolumn{1}{l}{$z_0$}
\end{tabular}
    \end{minipage}%
    \begin{minipage}{.5\linewidth}
      \centering
\begin{tabular}{cccc}
\multicolumn{4}{l}{$P_{2}(x)$=$x^{4}+x+1$}                                                                           \\
\multicolumn{1}{c|}{$s_3$} & \multicolumn{1}{c|}{$s_2$} & \multicolumn{1}{c|}{$s_1$} & $s_0$                     \\
\multicolumn{1}{c|}{0} & \multicolumn{1}{c|}{0}  & \multicolumn{1}{c|}{$s_4$} & $s_4$                     \\
\multicolumn{1}{c|}{0} & \multicolumn{1}{c|}{$s_5$}  & \multicolumn{1}{c|}{$s_5$}  & 0                     \\
\multicolumn{1}{c|}{$s_6$} & \multicolumn{1}{c|}{$s_6$} & \multicolumn{1}{c|}{0}  & 0                      \\ \hline
\multicolumn{1}{l|}{$z_3$} & \multicolumn{1}{l|}{$z_2$} & \multicolumn{1}{l|}{$z_1$} & \multicolumn{1}{l}{$z_0$}
\end{tabular}    \end{minipage} 
\caption{\small Two GF($2^4$) multiplications constructed using $P_{1}(x)$=$x^{4}+x^{3}+1$ and $P_{2}(x)$=$x^{4}+x+1$.}
\label{fig:4-bit-gf-multiplication}
\end{figure}
\begin{figure}[!htb]
\centering
\begin{tabular}{|c|l|}
\hline
output & \multicolumn{1}{c|}{polynomial expression} \\ \hline
$z_0$     &      (\textbf{$a_0$$b_0$})+$a_1$$b_3$+$a_2$$b_2$+$a_3$$b_1$                 \\ \hline
$z_1$     &      (\textbf{$a_0$$b_1$+$a_1$$b_0$})+$a_1$$b_3$+$a_2$$b_2$+$a_2$$b_3$+$a_3$$b_1$+$a_3$$b_2$           \\ \hline
$z_2$     &      (\textbf{$a_0$$b_2$+$a_1$$b_1$+$a_2$$b_0$})+$a_2$$b_3$+$a_3$$b_2$+$a_3$$b_3$                 \\ \hline
$z_3$     &      (\textbf{$a_0$$b_3$+$a_1$$b_2$+$a_2$$b_1$+$a_3$$b_0$})+$a_3$$b_3$                 \\ \hline
\end{tabular}
\caption{Extracted algebraic expressions of the four output bits of GF($2^4$) multiplier for $P(x)=x^4+x+1$.}
\label{fig:4-bit}
\end{figure}

\subsection{Irreducible Polynomials}\label{sec:irreducible_poly}

In general, there are various irreducible polynomials that can be used for a given field size, each resulting in a different multiplication result. For constructing efficient arithmetic functions over GF($2^m$), the irreducible polynomial is typically chosen to be a trinomial, $x^m$+$x^a$+1, or a pentanomial $x^m$+$x^a$+$x^b$+$x^c$+1 \cite{nist-recommend}. 
{For efficiency reason, coefficients $m,~a$ are chosen such that $m$ - $a$ $\geq$ $m/2$}. 

An example of constructing GF($2^4$) multiplication using two different irreducible polynomials is shown in Figure \ref{fig:4-bit-gf-multiplication}. We can see that each polynomial produces a unique multiplication result. The size of the corresponding multiplier can be estimated by counting the number of XOR operations in each multiplication. Since the number of AND and XOR operations for generating partial products (variables $s_{i}$ in Figure \ref{fig:4-bit-gf-multiplication}) is the same, the difference is only caused by the reduction of the corresponding polynomials modulo $P(x)$. The number of two-input XOR operations introduced by the reduction with $P(x)$ can be obtained as the number of terms in each column minus one. For example, the number of XORs using $P_1(x)$ is 3+1+2+3=9; and using $P_2(x)$, the number of XORs is 1+2+2+1=6.


As will be shown in the next section, given the structure of the GF($2^m$) multiplication, such as the one shown in Figure \ref{fig:4-bit-gf-multiplication}, one can readily identify the irreducible polynomial $P(x)$ used during the $GF$ reduction. This can be done by extracting the terms $s_k$ corresponding to the entry $s_m$ (here $s_4$) in the table and generating the irreducible polynomial beyond $x^m$. 
We know that $P(x)$ must contain $x_m$, and the remaining terms $x^k$ of $P(x)$ are obtained from the non-zero terms corresponding to the entry $s_m$. 
For example, for the irreducible polynomial $P_1(x)=x^4+x^3+x^0$, the terms $x^3$ and $x^0$ are obtained by noticing the placement of $s_4$ in columns $z_3$ and $z_0$. Similarly, for $P_2(x)=x^4+x^1+x^0$, the terms $x^1$ and $x^0$ are obtained by noticing that $s_4$ is placed in columns $z_1$ and $z_0$. 
The reason for it and the details of this procedure will be explained in the next section.


\section{Parallel Extraction in Galois Field} \label{sec:parallel_GF}

In this section, we introduce our method for extracting the unique algebraic expressions of the output bits (e.g. Figure \ref{fig:4-bit}) using computer algebraic method. This can be used to verify the GF($2^m$) multipliers when the binary encoding of inputs and output and the irreducible polynomial are given. We introduce a parallel function extraction framework in GF($2^m$), which allows us to individually extract the algebraic expression of each output bit. This framework is used for reverse engineering, since our reverse engineering approach is based on analyzing the algebraic expression of output bits in GF(2), as introduced in Section \ref{sec:introduction}. 

\subsection{Computer Algebraic model}
{
The circuit is modeled as a network of logic elements of arbitrary complexity, including basic logic gates (AND, OR, XOR, INV) and complex standard cell gates (AOI, OAI, etc.) generated by logic synthesis and technology mapping. We extend the algebraic model of Boolean operators developed in \cite{yu:2016-tcad-verification} for integer arithmetic to finite field arithmetic in $GF(2)$, i.e., modulo 2.} For example, the pseudo-Boolean model of XOR($a,b$)=$a+b$ $- 2ab$ is reduced to $(a + b + 2ab)$ mod $2$ = $(a + b)$ mod $2$. The following algebraic equations are used to describe basic logic gates in $GF(2^{m})$ \cite{kalla:tcad13}:

\vspace{-4mm}
\begin{equation}
     \begin{aligned}
      \text{~~} &\\
       & \neg a = 1 + a \\
       & a \wedge b = a\cdot b \\
       & a \vee b = a + b + a\cdot b \\
       & a \oplus b = a + b 
     \end{aligned}
\label{eq:boolean-poly}
\end{equation}

\subsection{Outline of the Approach}
{
Similarly to the work of \cite{ciesielski2015verification} and \cite{yu:2016-tcad-verification}, the arithmetic function computed by the circuits is obtained by transforming (rewriting) the polynomial representing the encoding of the primary outputs (called \textit{output signature}) into the polynomial at the primary inputs, the \textit{input signature}. The output signature of a $GF(2^{m})$ multiplier, $Sig_{out} = \sum _{i=0} ^{m-1} z_i x^i$, with $z_i \in GF(2)$. The input signature of a $GF(2^{m})$ multiplier, $Sig_{in}$ = $\sum _{i=0} ^{m-1} \mathbb{P}_i x^i$, with coefficients $\mathbb{P}_i \in GF(2)$ being product terms, and addition operation performed modulo 2. }
{If the irreducible polynomial $P(x)$ is provided, $Sig_{in}$ is know; otherwise, it will be computed by backward rewriting from $Sig_{out}$. The goal is to transform the output signature, $Sig_{out}$, using polynomial representation of the internal logic elements (\ref{eq:boolean-poly}), into an input signature $Sig_{in}$ in $GF(2^m)$, which determines the arithmetic function (specification) computed by the circuit.}

\textbf{Theorem 1:} \textit{Given a combinational arithmetic circuit in $GF(2^m)$, composed of logic gates, described by Eq. 1, input signature $Sig_{in}$ computed by backward rewriting is unique and correctly represents the function implemented by the circuit in $GF(2^m)$.}

\textbf{Proof:} The proof of correctness relies on the fact that each transformation step (rewriting iteration) is correct. That is, each internal signal is represented by an algebraic expression, which always evaluates to a {\it correct value} in $GF(2^{m})$. This is guaranteed by the correctness of the algebraic model in Eq. (\ref{eq:boolean-poly}), which can be proved easily by inspection. For example, the algebraic expression of \textit{XOR(a,b)} in $\mathbb{Z}_{2^m}$ is $a+b-2ab$. When implemented in $GF(2^{m})$, the coefficients in the expression must be in $GF(2)$, hence \textit{XOR(a,b)} in $GF{2^m}$ is represented by $a+b$. The proof of uniqueness is done by induction on $i$, the step of transforming polynomial $F_i$ into $F_{i+1}$. A detailed induction proof of this theorem is provided in \cite{ciesielski2015verification} for expressions in $\mathbb{Z}_{2^m}$. 

\hfill $\square$




\begin{algorithm}
\scriptsize
\caption{Backward Rewriting in $GF(2^{m})$}\label{alg:commonlogic}
\textbf{Input: Gate-level netlist of $GF(2^{m})$ multiplier}\\ 
\textbf{Input: Output signature $Sig_{out}$, and (optionally) input signature, $Sig_{in}$} \\
\textbf{Output: GF function of the design; return $Sig_{out}$==$Sig_{in}$}
\begin{algorithmic}[1]
\State $\mathcal{P}$=\{$p_{0},p_{1},...,p_{n}$\}: polynomials representing gate-level netlist
\State $F_{0}$=$Sig_{out}$
\For{each polynomial $p_{i}$ $\in \mathcal{P}$} 
\For{output variable $v$ of $p_{i}$ in $F_{i}$}
\State replace every variable $v$ in $F_{i}$ by algebraic expression of $p_{i}$
\State $F_{i}$ $\rightarrow$ $F_{i+1}$
\For{each monomial $M$ in $F_{i+1}$}
\If {the coefficient of $M$\%2==0 \\~~~~~~~~~~~~~~~~~or $M$ is a constant, $M$\%2==0}
\State remove $M$ from $F_{i+1}$
\EndIf
\EndFor
\EndFor
\EndFor \\
\Return $F_{n}$ and $F_{n}=?Sig_{in}$
\end{algorithmic}
\end{algorithm}

Theorem 1, together with the algebraic model of Boolean gates (\ref{eq:boolean-poly}), provide the basis for polynomial reduction using backward rewriting. This is described by Algorithm 1. The method takes the gate-level netlist of a GF($2^{m}$) multiplier as input and first converts each logic gate into an algebraic expression using Eq. (1). 
The rewriting process starts with the output signature $F_{0}=Sig_{out}$ and performs rewriting in reverse topological order, from outputs to inputs. It ends when all the variables in $F_{i}$ are primary inputs, at which point it becomes the input signature $Sig_{in}$ \cite{ciesielski2015verification}. 

{Each iteration includes two basic steps: 1) substitute the variable of the gate output using the expression in the inputs of the gate (Eq.1), and name the new expression $F_{i+1}$ (lines 3 - 6); and 2) simplify the new expression in two ways: a) by eliminating terms that cancel each other (as in the integer arithmetic case \cite{ciesielski2015verification}), and b) by removing all the monomials (including constants) that reduce to 0 in GF($2$) (line 3 and lines 7 - 10). The algorithm outputs the arithmetic function of the design in GF($2^m$) after $n$ iterations, where $n$ is the number of gates in the netlist. The final expression $F_{n}=Sig_{in}$ can be used to verify if the circuit performs the desired arithmetic function by checking if the computed polynomial $Sig_{in}$ matches the expected specification, if known. This equivalence check can be readily performed using canonical word-level representations, such as BMD \cite{bmd95} or TED \cite{ted:tcomp06} which can efficiently check equivalence of two polynomials. Alternatively, if the specification is not known, the computed signature can serve as the specification extracted from the circuit.}

\begin{figure}[t] 
\begin{center}
\includegraphics[scale=0.35]{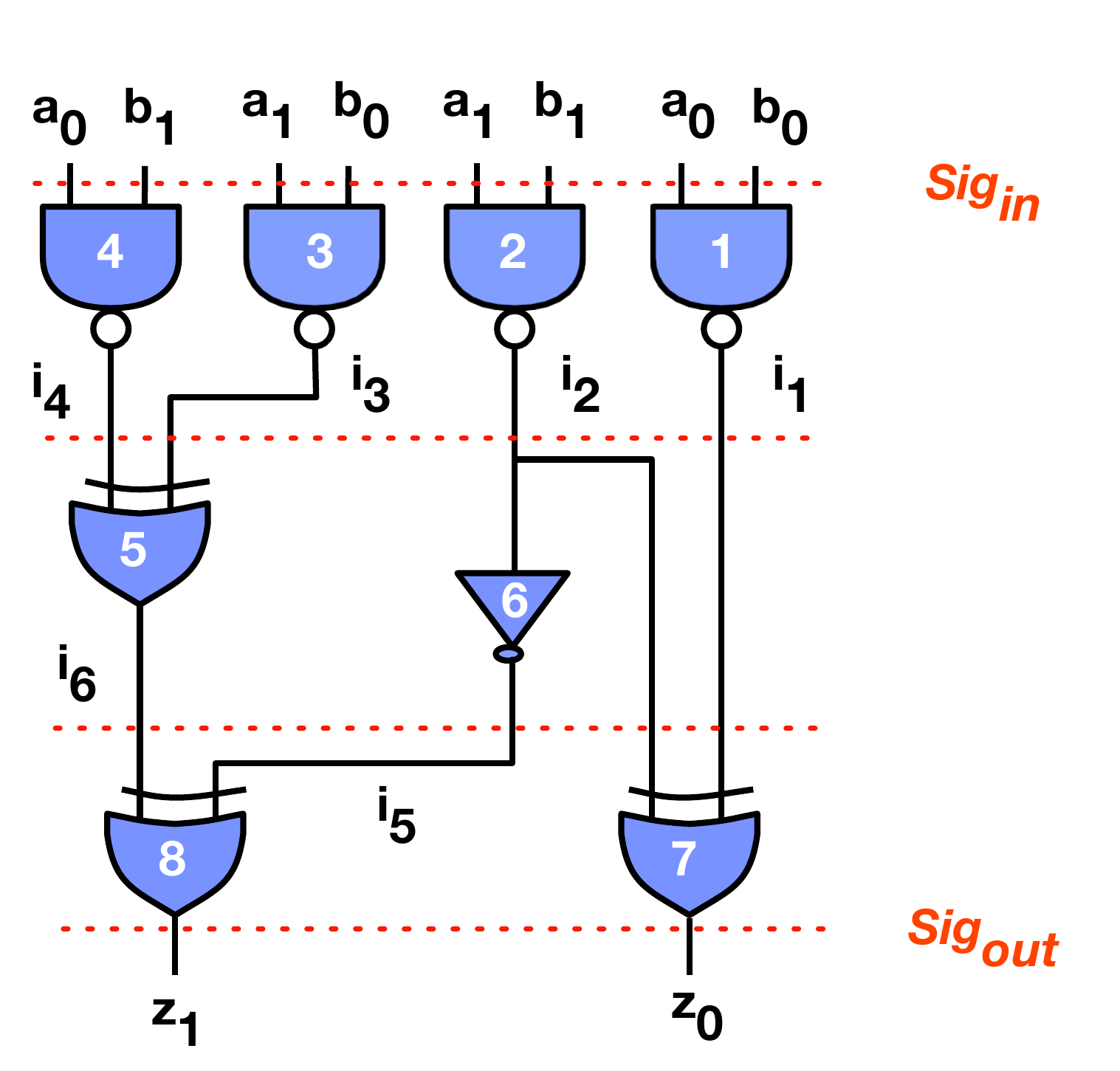}
\caption{The gate-level netlist of post-synthesized and mapped 2-bit multiplier over GF($2^2$). The irreducible polynomial is $P(x)=x^{2}+x+1$.}
\vspace{-2mm}
\label{fig:netlist-2bit}
\end{center}
\end{figure}

\begin{figure}[]
\centering
\small
\begin{tabular}{|l|c|}
\hline
$Sig_{out}$: $F_{init}$=$z_0$+x$z_1$                                                &  Eliminating terms           \\ \hline
G8: $F_8$=$z_0$+x($i_5$+$i_6$)                                             & \textit{-}  \\ \hline
G7: $F_7$=$i_1$+$i_2$+x($i_5$+$i_6$)                                           & \textit{-}  \\ \hline
G6: $F_6$=$i_1$+$i_2$+x($i_3$+$i_4$+$i_5$)                                     & \textit{-}  \\ \hline
G5: $F_5$=$i_1$+$i_2$+x($i_3$+$i_4$+$i_2$+1)                                   & \textit{-}  \\ \hline
G4: $F_4$=$i_1$+$i_2$+x($i_2$+$i_3$+$a_0$$b_1$)+2x                          & \textit{2x} \\ \hline
G3: $F_3$=$i_1$+$i_2$+x($i_2$+$a_1$$b_0$+$a_0$$b_1$+1)                      & \textit{-}  \\ \hline
G2: $F_2$=$i_1$+$a_1$$b_1$+1+x($a_1$$b_1$+$a_1$$b_0$+$a_0$$b_1$)+2x      & \textit{2x} \\ \hline
G1: $F_1$=$a_0$$b_0$+$a_1$$b_1$+2+x($a_1$$b_1$+$a_1$$b_0$+$a_0$$b_1$) & \textit{2}  \\ \hline
$Sig_{in}$: $a_0$$b_0$+$a_1$$b_1$+x($a_1$$b_1$+$a_1$$b_0$+$a_0$$b_1$) & \textit{-}  \\ \hline
\end{tabular}
\caption{Function extraction of a 2-bit $GF$ multiplier shown in Figure \ref{fig:netlist-2bit} using backward rewiring from PO to PI.}
\label{fig:rewriting}
\end{figure}

{\bf Example 2} (Figure \ref{fig:netlist-2bit}): We illustrate our method using a post-synthesized 2-bit multiplier in $GF(2^2)$, shown in Figure \ref{fig:netlist-2bit}. The irreducible polynomial is $P(x)$ = $x^{2}+x+1$. The output signature is $Sig_{out} = z_{0}$+$z_{1}x$, and input signature is $Sig_{in}=(a_{0}b_{0}$+$a_{1}b_{1}$)+($a_{1}b_{1}$+$a_{1}b_{0}$+$a_{0}b_{1}$)$x$. First, $F_{init}=Sig_{out}$ is transformed into $F_{8}$ using polynomial of gate $g8$, $z_{1}$=$i_{5}+i_{6}$ and simplified to $F_{8}=z_{0}+i_{5}x+i_{6}x$. Then, the polynomials $F_{i}$ are successively derived from $F_{i+1}$ and checked for a possible reduction. The first reduction happens when $F_{5}$ is transformed into $F_{4}$, where $i_{4}$ (at gate $g_4$) is replaced by ($1+a_{0}b_{0}$). After simplification, a monomial $2x$ is identified and removed by modulo 2 from $F_{4}$. Similar reductions are applied during the transformations $F_{3} \rightarrow F_{2}$ and $F_{2} \rightarrow F_{1}$. Finally, the function of the design is extracted as expression $F_{1}$. A complete rewriting process is shown in Figure \ref{fig:rewriting}. We can see that $F_{1}=Sig_{in}$, which indicates that the circuit indeed implements the $GF(2^2)$ multiplication with $P(x)$=$x^{2}+x+1$. 

An important observation is that the potential reductions take place only within the expression  associated with the same degree of polynomial ring ($Sig_{out}$). In other words, the reductions happen in a logic cone of every output bit $independently$ of other bits, regardless of logic sharing between the cones. For example, the reductions in $F_{4}$ and $F_{2}$ happen within the logic cone of output $z_{1}$ only. Similarly, in $F_{1}$, the reduction is within logic cone of $z_{0}$. Details of the proof are provided in \cite{yu:aspdac17}.



\subsection{Implementation} \label{sec:implementation}

This section describes the implementation of our parallel verification method for Galois field multipliers. Our approach takes the gate-level netlist as input, and outputs the extracted function of the design. It includes four steps: 

{\bf Step1: Convert netlist to equations.} Parse the gate-level netlist into algebraic equations based on Equation 1. The equations are listed in topological order, to be rewritten by backward rewriting in the next step. $m$ copies of this equation file will be made for a GF($2^m$) multiplier.

{\bf Step2: Generate signatures.} Split the output signature of GF($2^{m}$) multipliers into $m$ polynomials, with $Sig_{out\_i}$=$z_{i}$. Insert the new \textit{signatures} into the $m$ copies of the equation file generated from Step1. Each signature represents a single output bit.

\begin{figure}[!htb] 
\begin{center}
\includegraphics[scale=0.50]{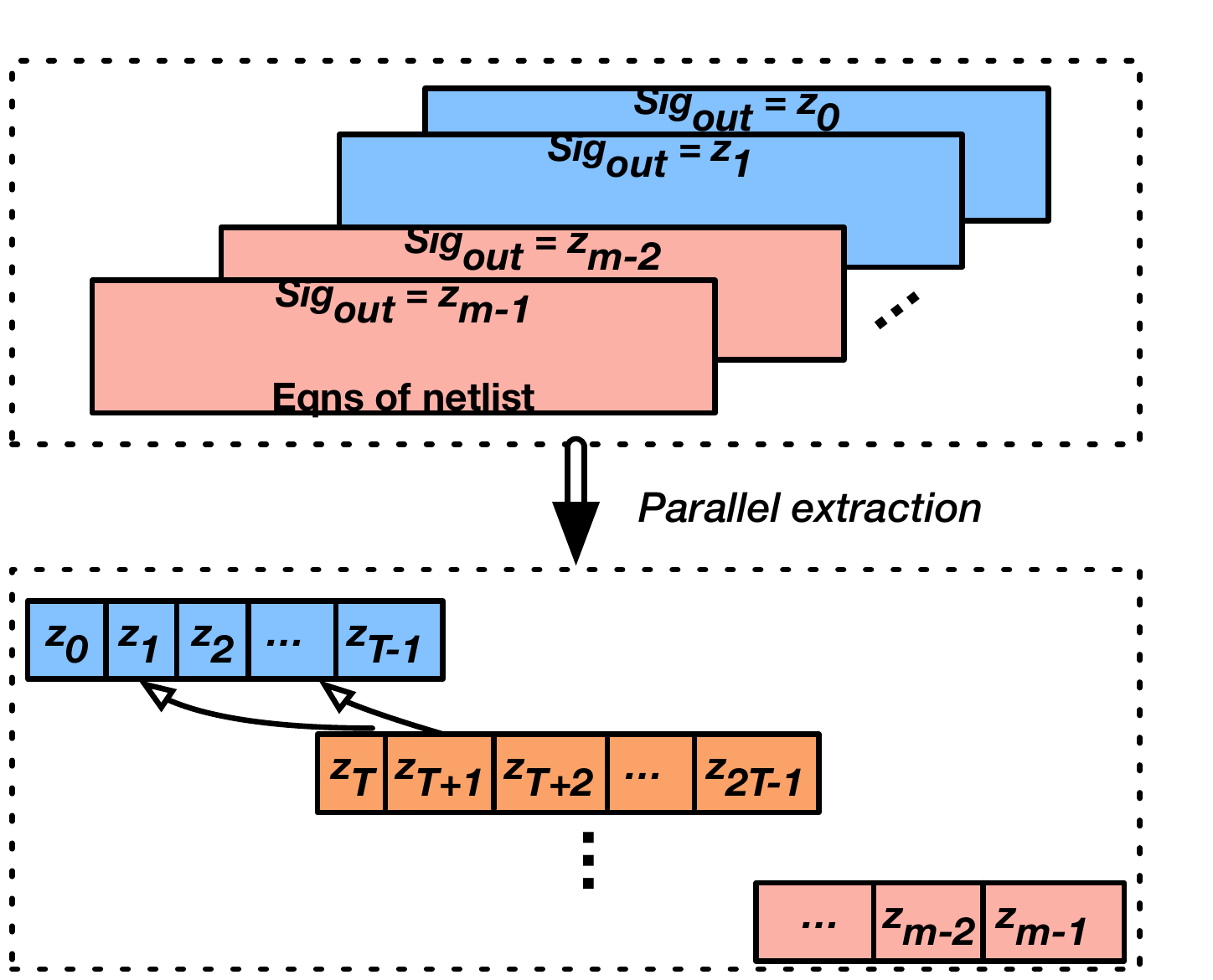}
\caption{Step3: parallel extraction of a GF($2^m$) multiplier with $T$ threads.}
\label{fig:flow}
\end{center}
\end{figure}

{\bf Step3: Parallel extraction.} Apply Algorithm 1 to each equation file to extract the polynomial expression of each output in parallel. In contrast to work on integer arithmetic \cite{ciesielski2015verification}, the internal expression of each output bit does not offer any polynomial reduction (\textit{monomial cancellations}) with other bits. 
Ideally, our approach can extract GF($2^m$) multiplier in $m$ threads. However, due to the limited computing resources, it is impossible to extract GF($2^m$) multipliers in $m$ threads when $m$ is very large. Hence, our approach puts a limit on the number of parallel threads $T$ (T = 5, 10, 20 and 30 have been tested in this work). This process is illustrated in Figure \ref{fig:flow}. The $m$ extraction tasks are organized into several task sets, ordered from LSB to MSB. In each set, the extractions are performed in parallel. Since the runtime of each extraction within the set can differ, the tasks in the next set will start as soon as any previous task terminated. 

{\bf Step4: Finalization.} Compute the final function of the multiplier. Once the algebraic expression of each output bit in GF($2$) is computed, our method computes the final function by constructing the $Sig_{out}$ using the rewriting process in step 3.

\noindent
{ Our algorithm uses a data structure that efficiently implements iterative substitution and elimination during backward rewriting. It is similar to the data structure employed in function extraction for integer arithmetic circuits  \cite{ciesielski2015verification}, suitably modified to support simplifications in finite fields algebra. Specifically, in addition to cancellation of terms with opposite signs, it performs modulo 2 reduction of monomials and constants.}
The data structure maintains the record of the terms (monomials) in the expression that contain the variable to be substituted. It reduces the cost of finding the terms that will have their coefficients changed during substitution. Each element represents one monomial consisting of the variables in the monomials and its coefficient. The expression data structure is a C++ object that represents a pseudo-Boolean expression, which contains of all the elements in the data structure. It supports both fast addition and fast substitution with two C++ maps, implemented as binary search trees, a terms map, and a substitution map. This data structure includes two cases of simplifications: 1) after substitution the coefficients of all the monomials are updated and the monomials with coefficient zero are eliminated; 2) the monomials whose coefficient modulo 2 evaluate to 0 are eliminated. The second case is applied after each substitution.

\begin{figure}[!htb]
\scriptsize
\centering
\begin{tabular}{|l|c|l|c|}
\hline
$Sig_{out0}$=$z_0$ & \begin{tabular}[c]{@{}c@{}}elim\end{tabular} & $Sig_{out1}$=x$\cdot$$z_1$ & \begin{tabular}[c]{@{}c@{}}elim\end{tabular} \\ \hline
G8: $z_0$ & \textit{-} & G8: $i_5$$x$+$i_6$$x$ & - \\ \hline
G7: $i_1$+$i_2$ & \textit{-} & G7: $i_5$$x$+$i_6$$x$ & - \\ \hline
G6: $i_1$+$i_2$ & \textit{-} & G6: $i_2$$x$+x+$i_6$$x$ & - \\ \hline
G5: $i_1$+$i_2$ & \textit{-} & G5: $i_2$$x$+x+$i_3$$x$+$i_4$$x$ & - \\ \hline
G4: $i_1$+$i_2$ & \textit{-} & G4: $i_2$$x$+\textbf{x}+$i_3$$x$+$a_0$$b_1$$x$+\textbf{x} & 2x \\ \hline
G3: $i_1$+$i_2$ & \textit{-} & G3: $i_2$$x$+$a_1$$b_0$$x$+x+$a_0$$b_1$$x$ & - \\ \hline
G2: $i_1$+$a_1$$b_1$+1 & \textit{-} & G2: $a_1$$b_1$$x$+\textbf{x}+$a_1$$b_0$$x$+\textbf{x}+$a_0$$b_1$$x$ & 2x \\ \hline
G1: \textbf{1}+$a_0$$b_0$+$a_1$$b_1$+\textbf{1} & \textit{2} & G1: x($a_1$$b_1$+$a_1$$b_0$+$a_0$$b_1$) & - \\ \hline
\multicolumn{4}{|l|}{$z_0$=$a_0$$b_0$+$a_1$$b_1$, $z_1$=x($a_1$$b_1$+$a_1$$b_0$+$a_0$$b_1$)} \\ \hline
\end{tabular}
\caption{\small Extracting the algebraic expression of $z_0$ and $z_{1}$ in Fig. \ref{fig:rewriting}.}
\label{fig:parallel}
\end{figure}

{\bf Example 3} (Figure \ref{fig:parallel}): We illustrate our parallel extraction method using a 2-bit multiplier in GF($2^2$) in Figure \ref{fig:netlist-2bit}. The output signature $Sig_{out}$ = $z_0$+$z_{1}x$ is split into two signatures, $Sig_{out0}=z_0$ and $Sig_{out1}=z_1$. Then, the rewriting process is applied to $Sig_{out0}$ and $Sig_{out1}$ in parallel. When $Sig_{out0}$ and $Sig_{out1}$ have been successfully extracted, the two signatures are merged into $Sig_{out0}$ + $x \cdot $$Sig_{out1}$, resulting in the polynomial $Sig_{in}$. In Figure \ref{fig:rewriting}, we can see that elimination happens three times ($F_4$, $F_2$, and $F_1$). As expected, this happens within each element in GF($2^n$). In Figure \ref{fig:parallel} one elimination in $Sig_{out0}$ and two eliminations in $Sig_{out1}$ have been done independently, as shown earlier (refer to Example 2).

\section{Reverse Engineering} \label{sec:reverse-engin}

In this section, we present our approach to perform reverse engineering of GF($2^m$) multipliers. Using the \textit{extraction} technique presented in the previous section, we can extract the algebraic expression of each output bit. In contrast to the algebraic techniques of \cite{PrussKE16TCAD}\cite{yu:2016-tcad-verification}, our extraction technique can extract the algebraic expression of each output bit independently. This means that the extraction can be done without the knowledge of the bit position of the inputs and outputs. Two theorems are provided and proved to support this claim. 

In a GF($2^m$) multiplication, let $s_{i}$ ($i$ $\in$ \{0,1,...,2$m$-1\}) be a set of partial products generated by AND gates and combined with an XOR operations. For example, in Figure \ref{fig:4-bit-gf-multiplication}, there are six product sets, $s_0$, $s_1$, ..., $s_6$, where $s_1$=$a_1$$b_0$+$a_0$$b_1$; or written as a set: $s_1$=\{$a_1$$b_0$, $a_0$$b_1$\}, etc. These product sets are divided into two groups: 
those with index $i \le m-1$, called \textit{in-field} product sets; and those with index $i \ge m$, called \textit{out-of-field} product sets. The in-field product sets $s_i$, in this case  $s_0, s_1, s_2, s_3$, correspond to the output bits $z_i$. The out-of-field product sets will be reduced into the field GF($2^m$) using mod $P(x)$ operation, and assigned to the respective output bit, as determined by $P(x)$. In the case of Figure \ref{fig:4-bit-gf-multiplication}, the out-of-field sets are $s_4$, $s_5$, $s_6$.
In general, for a GF($2^m$) multiplication, $m$ product sets are \textit{in-field}, and $m$-1 product sets are \textit{out-of-field} \cite{yu2017reverse}.  

\subsection{Output Encoding Determination}

We will now demonstrate how to determine the encoding, and hence bit position, of the outputs.

{
\textbf{Theorem 2:} Given a GF($2^m$) multiplication, the in-field product sets ($s_0$, $s_1$, ..., $s_{m-1}$) appear in exactly one element of GF($2^m$) each, and the out-of-field product sets ($s_m$, $s_{m+1}$, ..., $s_{2m-1}$) appear in at least two elements (outputs) of GF($2^m$), as a result of reduction mod $P(x)$.

\textbf{Proof:} 
An irreducible polynomial in GF($2^m$) has the standard form $P(x) = x^m + P'(x)$, where the tail polynomial $P'(x)$ contains at least two monomials $x^d$ with degree $d < m$. For example, there are two such monomials for a trinomial, four for pentanomial, etc. 
Since $P(x) = 0$ we have $x^m = P'(x)$ in GF($2^m$). Hence the variable $x^m$, associated with the first out-of-field partial product set $s^m$ will appear in at least two outputs, determined by $P'(x)$. 
Other variables, $x^k$, associated with out-of-field partial product set $s_k$, for $k > m$, can be expressed as $x^k = x^{k-m} x^m = x^{k-m} P'(x)$ and will contain at least two elements.
 QED
\hfill $\square$
}

{
In fact, the number of outputs in which the out-of-field set $s_k$ will appear is equal to the number of monomials in the above product $x^{k-m} P'(x)$, provided that every monomial $x^j$ with $j > m$ is recursively reduced mod $P'(x)$, i.e., by using relation $x^m = P'(x)$.
We illustrate this fact with an example of multiplication in GF($2^4$) using irreducible polynomial $P_1(x)= x^4+x^3+1$ shown in the left side of Figure \ref{fig:4-bit-gf-multiplication}. The in-field sets, associated with outputs $z_0, z_1, z_2, z_3$, are $s_0, s_1, s_2, s_3$. Since $P_1(x) = x^4+x^3+1 = 0$, we obtain $ x^4 = x^3+1$. This means that set $s_4$ appears in two output columns, $z_3$ and $z_0$. Then
\[ x^5 = x\cdot x^4 = x(x^3+1) = x^4 + x = x^3+x+1,\]
which means that $s_5$ appears in three outputs: $z_3, z_1, z_0$. Finally,
\[ x^6 = x\cdot x^5 = x(x^3+x+1) = x^4 + x^2+x = x^3+x^2+x+1,\]
that is, $s_6$ will appear in four outputs: $z_3, z_2, z_1, z_0$. 
As expected, this matches the left Table in Figure \ref{fig:4-bit-gf-multiplication}.
Note the recursive derivation of $x^k$ for $k > m$, which increases the number of columns to which a given set $s_k$ is assigned.}

Based on Theorem 2, we can find the in-field product sets, $s_0$, $s_1$, ..., $s_{m-1}$, by searching the unique products in the resulting algebraic expressions of the output bits. In this context, \textit{unique products} are the products that exist in only one of the extracted algebraic expressions. Since the in-field product set indicates the bit position of the output, we can determine the bit positions of the output bits as soon as all the in-field product sets are identified.

\textbf{Example 4 (Figure \ref{fig:4-bit}):} We illustrate the procedure of determining bit positions with an example of a GF($2^4$) multiplier implemented using irreducible polynomial $P_2(x)$=$x^4$+$x$+$1$ (see Figure \ref{fig:4-bit-gf-multiplication}). Note that in this process the labels do not offer any knowledge of the bit positions of inputs and outputs. The extracted algebraic expressions of the four output bits are shown in Figure \ref{fig:4-bit}. The labels of the variables do not indicate any binary encoding information. We first identify the unique products that include set $s_0$=$a_0$$b_0$ in algebraic expression of $z_{0}$; set $s_1$=($a_0$$b_1$+$a_1$$b_0$) in $z_{1}$; set $s_2$=($a_0$$b_2$+$a_1$$b_1$+$a_2$$b_0$) in $z_{2}$; and set $s_3$=($a_0$$b_3$+$a_1$$b_2$+$a_2$$b_1$+$a_3$$b_0$) in $z_{3}$. Note that the number of products in the in-field product set $s_{i}$ is $i$. Hence, we find all the in-field product sets and their relation to the extracted algebraic to be as follows: 

\vspace{2mm}
\noindent
$s_0$ = $a_0$$b_0$, $z_0$ $\rightarrow$ Least significant bit (LSB)

\noindent
$s_1$ = $a_0$$b_1$+$a_1$$b_0$, $z_1$ $\rightarrow$ $2^{nd}$ output bit

\noindent
$s_2$ = $a_0$$b_2$+$a_1$$b_1$+$a_2$$b_0$, $z_2$ $\rightarrow$ $3^{rd}$ output bit

\noindent
$s_3$ = $a_0$$b_3$+$a_1$$b_2$+$a_2$$b_1$+$a_3$$b_0$, $z_3$ $\rightarrow$ Most significant bit (MSB)

\subsection{Input Encoding Determination}

\begin{algorithm}
\scriptsize
\caption{Input encoding determination for $GF(2^{m})$}\label{alg:commonlogic}
\textbf{Input: a set of algebraic expressions represent the in-field product sets $S$}\\ 
\textbf{Output: bit position of input variables}
\begin{algorithmic}[1]
\State $S$=\{$s_0, s_1, ..., s_{m-1}$\}
\State initialize a vector of variables $V$ $\leftarrow$ \{\}
\For{i=0, i$\leq$m-1, i++} 
\For{each variable $v$ in algebraic expression of $s_{i}$}
\If {$v$ does not exist in $V$}
\State assign bit position value of $v$ = $i$
\State store $v$ in variable set $V$
\EndIf
\EndFor
\EndFor \\
\Return $V$
\end{algorithmic}
\end{algorithm}

We can now determine the bit position of the input variables using the procedure outlined in Algorithm 2. The input bit position can be determined by analyzing the in-field product sets, obtained in the previous step. Based on the GF multiplication algorithm, we know that $s_0$ is generated by an AND function with two LSBs of the two inputs; and the two products in $s_1$ are generated by the AND and XOR operations using two LSBs and two $2^{nd}$ input bits, etc. For example in a GF($2^4$) multiplication (Figure \ref{fig:4-bit-gf-multiplication}), $s_0$=$a_0$$b_0$, where $a_0$ and $b_0$ are LSBs; $s_1$=$a_1$$b_0$+$a_0$$b_1$, where $a_0$, $b_0$ are LSBs; $a_1$, $b_1$ are $2^{nd}$ LSBs. This allows us to determine the bit position of the input bits recursively by analyzing the algebraic expression of $s_{i}$. We illustrate this with the GF($2^4$) multiplier implemented using $P_2(x)$ = $x^4$+$x$+$1$ (Figure \ref{fig:4-bit}).

\textbf{Example 5 (Algorithm 2):} The input of our algorithm is a set of algebraic expressions of the in-field product sets, $s_{0}$, $s_{1}$, $s_{2}$, $s_{3}$ (line 1). We initialize vector $V$ to store the variables in which their bit positions are assigned (line 2). The first algebraic expression is $s_0$. Since the two variables, $a_0$ and $b_0$ are not in $V$, the bit positions of these two variables are assigned index $i=0$ (line 4-8). In the second iteration, $V$=\{$a_0$, $b_0$\}, and the input algebraic expression is $s_1$, including variables $a_0$, $b_0$, $a_1$ and $b_1$. Because $a_1$ and $b_1$ are not in $V$, their bit position is $i=1$. The loop ends when all the algebraic expressions in $S$ have been visited, and returns $V$=\{$(a_0, b_0)_{0}$, $(a_1, b_1)_{1}$, $(a_2, b_2)_{2}$, $(a_3, b_3)_{3}$\}. The subscripts are the bit position values of the variables returned by the algorithm. Note that this procedure only gives the bit position of the input bits; the information of how the input words are constructed is unknown. There are $2^{m-1}$ combinations from which the words can be constructed using the information returned in $V$. For example, the two input words can be $W_{0}$=$a_0$+2$a_1$+4$b_2$+8$a_3$ and $W_{1}$=$b_0$+2$b_1$+4$a_2$+8$b_3$; or they can be $W'_{0}$=$a_0$+2$a_1$+4$b_2$+8$b_3$ and $W'_{1}$=$b_0$+2$b_1$+4$a_2$+8$a_3$. Although there may be many combinations for constructing the input words, the specification of the GF($2^m$) is unique.

\subsection{Extraction of the Irreducible Polynomial}
\textbf{Theorem 3:} \textit{Given a multiplication in GF($2^m$), let the first out-of-field product set be $s_{m}$. Then, the irreducible polynomial $P(x)$ includes monomials $x^m$ and $\{x^i\}$ iff all products in the set $s_{m}$ appear in the algebraic expression of the $i^{th}$ output bits, for all $i$ $ < $ $m$.}

\textbf{Proof:} 
Based on the definition of field arithmetic for GF($2^m$), the polynomial basis representation of $s_{m}$ is $x^m s_m$. To reduce $s_{m}$ into elements in the range [0, $m-1$], the field reductions are performed modulo irreducible polynomial $P(x)$ with highest degree $m$ ($c.f.$ the proof of Theorem 2). 
As before, let $P(x)$ = $x^m + P'(x)$. Then, 
\[ x^m s_m ~mod~ (x^m+P'(x))~=~s_{m}P'(x) \]
{Hence, if $x^i$ exists in $P'(x)$, it also exists in $P(x)$. Therefore, $x^i$ exists in $P(x)$, iff $x^i s_m $ exists in $x^m s_m$ mod $P(x)$.}

\hfill $\square$

Even though the input bit positions have been determined in the previous step, we cannot directly generate $s_m$ since the combination of the input bits for constructing the input words is still unknown. In Example 5 ($m$=4), we can see that $s_m$=\{$a_1$$b_3$, $a_2$$b_2$, $a_3$$b_1$\} when input words are $W_{0}$ and $W_{1}$; but $s_m$=\{$a_1$$a_3$, $a_2$$b_2$, $b_1$$b_3$\} when inputs words are $W_{0}'$ and $W'_1$. To overcome this limitation, we create a set of products $s_m'$, which includes all the possible products that can be generated based on all input combinations. The set $s_m'$ includes the $true$ products, i.e., those that exist in the first out-of-field product set; and it also includes some $dummy$ products. The dummy products are those that never appear in the resulting algebraic expressions. Hence, we first generate the set $s_m'$ and eliminate the dummy products by searching the algebraic expressions. After this, we obtain $s_m$. Then, we use $s_m$ to extract the irreducible polynomial $P(x)$ using Algorithm 3. 

{\bf Example 6: }We illustrate the method of reverse engineering the irreducible polynomial using the $GF(2^4)$ multiplier of Fig. 1. The procedure is outlined in Algorithm 3. The extracted algebraic expressions $S$ (line 1 at Algorithm 3) is shown in Figure \ref{fig:4-bit}. The bit position of input bits is determined by Algorithm 2 (line 2). Based on the result of Algorithm 2, we generate $s_m'$=\{$a_1$$a_3$, $b_1$$b_3$, $a_2$$b_2$, $a_3$$b_1$, $a_1$$b_3$\}. To eliminate the dummy products from $s_m'$, we search all algebraic expressions in $S$, and eliminate the products that cannot be part of the resulting products. In this case, we find that $a_1$$a_3$ and $b_1$$b_3$ are the dummy products. Hence, we get $s_m$=\{$a_3$$b_1$, $a_2$$b_2$, $a_1$$b_3$\} (line 3). Based on the definition of irreducible polynomial, $P(x)$ must include $x^m$; in this example $m=4$ (line 4). While looping over all the algebraic expressions, the expressions for $z_{0}$ and $z_{1}$ contain all the products of $s_m$. Hence, $x^0$ and $x^1$ are included in $P(x)$, so that $P(x)$ = $x^4$+$x^1$+$x^0$. We can see that it is the same as $P_2(x)$ in Figure \ref{fig:4-bit-gf-multiplication}.

\begin{algorithm}
\scriptsize
\caption{Extracting irreducible polynomial in $GF(2^{m})$}\label{alg:commonlogic}
\textbf{Input: the algebraic expressions of output bits $S$}\\ 
\textbf{Input: the first out-of-field product set $s_m$}\\ 
\textbf{Output: Irreducible polynomial $P(x)$}
\begin{algorithmic}[1]
\State $S$ = \{$exp_0, exp_1, ..., exp_{m-1}$\}
\State $V$ $\leftarrow$ Algorithm 2($S$)
\State $s_m$ $\leftarrow$ $eliminate$\_$dummy$($s_m'$ $\leftarrow$ $V$, $S$)
\State $P(x)$=$x^m$: initialize irreducible polynomial
\For{i=0, i$\leq$m-1, i++} 
\If {all products in $s_{m}$ exist in $exp_i$}
\State $P(x)$~+=~$x^i$
\EndIf
\EndFor \\
\Return $P(x)$
\end{algorithmic}
\end{algorithm}

\begin{table*}[]
\scriptsize
\centering
\begin{tabular}{|r|r|r|r|r|r|r|r|r|r|}
\hline
\multicolumn{2}{|c|}{\textit{Mastrovito}} & \multicolumn{2}{c|}{\cite{kalla:dac2014}} & \multicolumn{6}{c|}{This work} \\ \hline
\multirow{2}{*}{Op size} & \multirow{2}{*}{\# equations} & \multirow{2}{*}{\begin{tabular}[c]{@{}c@{}}Runtime\\ (sec)\end{tabular}} & \multirow{2}{*}{\begin{tabular}[c]{@{}c@{}}Mem\\ (MB)\end{tabular}} & \multicolumn{5}{c|}{Runtime (s)} & Mem* \\ \cline{5-9} 
 &  &  &  & \textit{T=1} & \textit{T=5} &  \textit{T=10} & \textit{T=20} & \textit{T=30} & \textit{T=1*} \\ \hline
32 & 5,482 & 1 & 3 & 5 & 2 & 1 & 1 & 1 & 10 MB \\ \hline
48 & 12,228 & 8 & 13 & 9 & 6 & 3 & 3 & 2 & 21 MB \\ \hline
64 & 21,814 & 29 & 21 & 19 & 11 & 8 & 7 & 7 & 37 MB \\ \hline
96 & 51,412 & 195 & 45 & 68 & 38 & 26 & 20 & 23 & 84 MB \\ \hline
128 & 93,996 & 924 & 91 & 153 & 91 & 63 & 55 & 57 & 152 MB \\ \hline
163 & 153,245 & 3546 & 161 & 336 & 192 & 137 & 121 & 113 & 248 MB \\ \hline
233 & 167,803 & 4933 & 168 & 499 & 294 & 212 & 180 & 171 & 270 MB \\ \hline
283 & 399,688 & 30358 & 380 & 1580 & 890 & 606 & 550 & 530 & 642 MB \\ \hline
571 & 1628,170 & \textit{TO} & - & 13176 & 7980 & 5038 &  \textit{MO} & \textit{MO} & 2.6 GB \\ \hline
\end{tabular}
\caption{Results of verifying Mastrovito multipliers using our parallel approach. $T$ is the number of threads. $MO$=Memory out of 32 GB. $TO$=Time out of 18 hours. \\(*\textit{T=1} shows the maximum memory usage of a single thread.)}
\label{tbl:mas-verification}
\end{table*}
\begin{table*}[]
\scriptsize
\centering
\begin{tabular}{|r|r|r|r|r|r|r|r|r|r|}
\hline
\multicolumn{2}{|c|}{\textit{Montgomery}} & \multicolumn{2}{c|}{\cite{kalla:dac2014}} & \multicolumn{6}{c|}{This work} \\ \hline
\multirow{2}{*}{Op size} & \multirow{2}{*}{\# equations} & \multirow{2}{*}{\begin{tabular}[c]{@{}c@{}}Runtime\\ (sec)\end{tabular}} & \multirow{2}{*}{\begin{tabular}[c]{@{}c@{}}Mem\\ (MB)\end{tabular}} & \multicolumn{5}{c|}{Runtime (s)} & Mem* \\ \cline{5-9} 
 &  &  &  & \textit{T=1} & \textit{T=5} &  \textit{T=10} & \textit{T=20} & \textit{T=30} & \textit{T=1*} \\ \hline
32 & 4,352 & 2 & 3 & 5 & 3 & 2 & 1 & 2 & 8 MB \\ \hline
48 & 9,602 & 14 & 13 & 34 & 18 & 11 & 9 & 6 & 16 MB \\ \hline
64 & 16.898 & 63 & 21 & 80 & 45 & 31 & 28 & 27 & 27 MB \\ \hline
96 & 37,634 & 554 & 45 & 414 & 234 & 157 & 133 & 142 & 59 MB \\ \hline
128 & 66,562 & 1924 & 68 & 335 & 209 & 121 & 115 & 110 & 95 MB \\ \hline
163 & 107,582 & 12063 & 101 & 2505 & 1616 & 1172 & 1095 & 1008 & 161 MB \\ \hline
233 & 219,022 & \textit{TO} & 168 & 1240 & 722 & 565 & 457 & 480 & 301 MB \\ \hline
283 & 322,622 & \textit{TO} & 380  &  32180 & 19745 & 17640 & 15300 & 14820 & 488 MB \\ \hline
\end{tabular}
\caption{Results of verifying \textit{Montgomery} multipliers using our parallel approach. $T$ is the number of threads. $TO$=Time out of 18 hours.\\(*\textit{T=1} shows the maximum memory usage of a single thread.) }
\label{tbl:mont-verification}
\end{table*}


In summary, using the framework presented in Section \ref{sec:implementation}, we first extract the algebraic expressions of all output bits. Then, we analyze the algebraic expressions to find the bit position of the input bits and the output bits, and extract the irreducible polynomial $P(x)$. In the example of the GF($2^4$) multiplier implemented using $P(x)$ = $x^4$+$x$+$1$, shown in Figure \ref{fig:4-bit-gf-multiplication}, the final results returned by our approach gives the following: 1) the input bits set $V$= \{$(a_0, b_0)_{0}$, $(a_1, b_1)_{1}$, $(a_2, b_2)_{2}$, $(a_3, b_3)_{3}$\}, where the subscripts represent the bit position; 2) $z_0$ is the least significant bit (LSB), $z_1$ is the $2^{nd}$ output bit, $z_2$ is the $3^{rd}$ output bit, and $z_3$ is the most significant bit (MSB); 3) irreducible polynomial is $P(x)$ = $x^4$+$x$+$1$; 4) the specification can be verified using the approach presented in Section \ref{sec:parallel_GF} with the reverse engineered information.

\section{Results} \label{sec:results}

The experimental results of our method are presented in two subsections: 
1) evaluation of parallel verification of GF($2^m$) multipliers; and 
2) evaluation of reverse engineering of GF($2^m$) multipliers.
The results given in this section include data (total time and maximum memory) for the entire verification or reverse engineering process, including translating the gate-level verilog netlist to the algebraic equation, performing backward rewriting and other required functions.

\subsection{Parallel Verification of GF($2^m$) Multipliers}

The verification technique for GF($2^m$) multipliers presented in Section \ref{sec:parallel_GF} was implemented in C++. It performs backward rewriting with variable substitution and polynomial reductions in Galois field in parallel fashion.
The program was tested on a number of combinational gate-level $GF(2^{m})$ multipliers taken from \cite{PrussKE16TCAD}, including the Montgomery multipliers \cite{koc1998montgomery} and Mastrovito multipliers \cite{sunar1999mastrovito}. The bit-width of the multipliers varies from 32 to 571 bits. The verification results for various Galois field multipliers obtained using SAT, SMT, ABC \cite{abc-link}, and Singular \cite{singular}, have already been presented in works of \cite{kalla:tcad13} and \cite{PrussKE16TCAD}. They clearly demonstrate that techniques based on computer algebra perform significantly better than other known techniques. Hence, in this work, we only compare our approach to those two, and specifically to the tool described in \cite{PrussKE16TCAD}. However, in contrast to the previous work on Galois field verification,
all the GF($2^m$) multipliers used in this paper are bit-blasted gate-level implementations. The bit-level multipliers are taken from \cite{PrussKE16TCAD} and mapped onto gate-level circuits using ABC \cite{abc-link}. 
Our experiments were conducted on a PC with Intel(R) Xeon CPU E5-2420 v2 2.20 GHz $\times$12 with 32 GB memory. As described in the next section, our technique can verify Galois field multipliers in multiple threads by applying Algorithm 1 to each output bit in parallel. The number of threads is given as input to the tool. 

The experimental results of our approach and comparison with \cite{PrussKE16TCAD} are shown in Table \ref{tbl:mas-verification} for gate-level Mastrovito multipliers with bit-width varying from 32 to 571 bits. These multipliers are directly mapped using ABC without any optimization. The largest circuit includes over 1.6 million gates. This is also the number of polynomial equations and the number of rewriting iterations (see Section \ref{sec:parallel_GF}). The results generated by the tool, presented in \cite{PrussKE16TCAD} are shown in columns 3 and 4 of Table \ref{tbl:mas-verification}. We performed four different series of experiments, with the number of threads $T$ varying from 5 to 30. The table shows CPU runtime and memory usage for different values of $T$. The timeout limit (TO) was set to 12 hours and memory limit (MO) to 32 GB. The experimental results show that our approach provides on average 26.2$\times$, 37.8$\times$, 42.7$\times$, and 44.3$\times$ speedup, for $T=$ 5, 10, 20, and 30 threads, respectively. Our approach can verify the multipliers up to 571 bit-wide multipliers in 1.5 hours, while that of \cite{PrussKE16TCAD} fails after 12 hours.

The reported memory usage of our approach is the maximum memory usage {\it per thread}. This means that our tool experiences maximum memory usage with all $T$ threads running in the process; in this case, the memory usage is $T \cdot Mem$. This is why the 571-bit Mastrovito multipliers could be successfully verified with $T$ = 5 and 10, but failed with $T$ = 20 and 30 threads. For example, the peak memory usage of 571-bit Mastrovito multiplier with $T=20$ is $2.6 \times 20=52$ GB, which exceeds the available memory limit.

We also tested Montgomery multipliers with bit-width varying from 32 to 283 bits; the results are shown in Table \ref{tbl:mont-verification}. These experiments are different than those in \cite{PrussKE16TCAD}. In our work, 
we first flatten the Montgomery multipliers before applying our verification technique. 
That is, we assume that only the positions of the primary inputs and outputs are known, without the knowledge of any high-level structure. In contrast, \cite{PrussKE16TCAD} verifies the Montgomery multipliers that are represented with four hierarchical blocks. For 32- to 163-bit Montgomery multipliers, our approach provides on average a 9.2$\times$, 15.9$\times$, 16.6$\times$, and 17.4$\times$ speedup, for $T=$ 5, 10, 20, and 30, respectively. Notice that \cite{PrussKE16TCAD} cannot verify the flattened Montgomery multipliers larger than 233 bits in 12 hours. 

Analyzing Table \ref{tbl:mas-verification} we observe that the rewriting technique of our approach when applied to Montgomery multipliers require significantly more time than for Mastrovito multipliers. The main reason for this difference is the internal architecture of the two multiplier types. Mastrovito multipliers are obtained directly from the standard multiplication structure, with the partial product generator followed by an XOR-tree structure, as in modular arithmetic. 
Since the algebraic model of XOR in GF arithmetic is linear, the size of the polynomial expressions generated during rewriting of this architecture is relatively small. In contrast, in a Montgomery multiplier the two inputs are first transformed into Montgomery form; the products of these Montgomery forms are called \textit{Montgomery products}. 
Since the polynomial expressions in Montgomery forms are much larger than partial products, the increase in size of intermediate expressions is unavoidable.

\subsubsection{\textbf{Dependence on $P(x)$}}

In Table \ref{tbl:mont-verification}, we observe that CPU runtime for verifying a 163-bit multiplier is greater than that of a 233-bit multiplier. This is because the computational complexity depends not only on the bit-width of the multiplier, but also on the irreducible polynomial $P(x)$ used in constructing the multiplier. 

We illustrate this fact using two GF($2^4$) multiplications implemented using two different irreducible polynomials (c.f. Figure \ref{fig:4-bit-gf-multiplication}). We can see that for $P_1(x)$=$x^{4}+x^{3}+1$, the longest logic paths for $z_{3}$ and $z_{0}$, include ten and seven products that need to be generated using XORs, respectively. However, when $P_2(x)$=$x^{4}+x+1$, the two longest paths, $z_{1}$ and $z_{2}$, have only seven and six products. This means that the GF($2^4$) multiplication requires 9 XOR operations using $P_{1}(x)$ and 6 XOR operations using $P_{2}(x)$. In other words, the gate-level implementation of the multiplier implemented using $P_{1}(x)$ has more gates compared to $P_{2}(x)$. In conclusion, we can see that irreducible polynomial $P(x)$ has significant impact on both design cost and the verification time of the GF($2^m$) multipliers. 



\subsubsection{\textbf{Runtime vs. Memory}}\label{sec:parallel_analysis}

\begin{figure}[!hbt]
  \centering

        \includegraphics[width=0.42\textwidth]{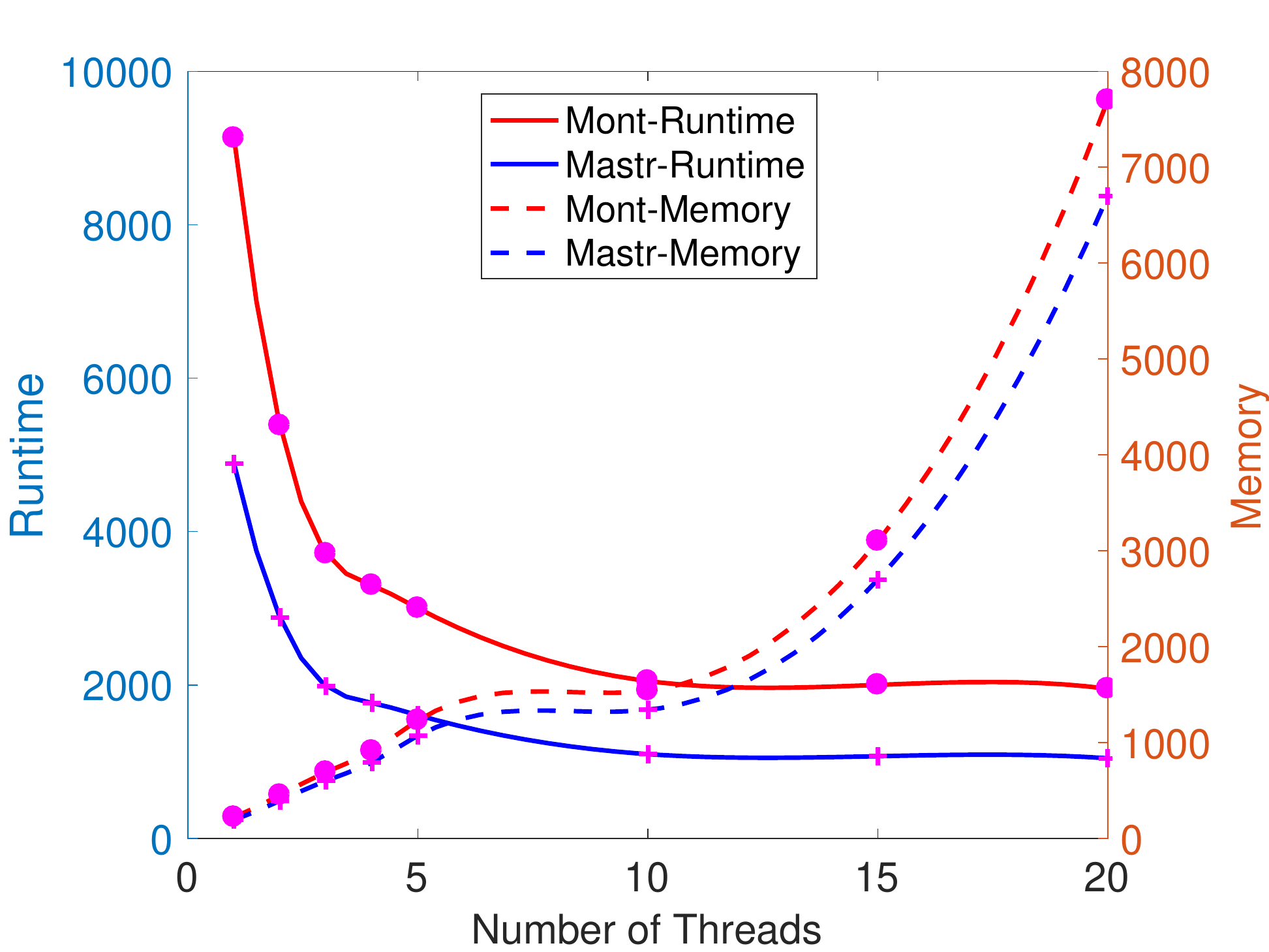}
\caption{Runtime and memory usage of parallel verification approach as a function of the number of threads $T$.}
\label{fig:tradeoff}
\end{figure}

In this section, we discuss the tradeoff of runtime and memory usage of our parallel approach to Galois Field multiplier verification. The plots in Figure \ref{fig:tradeoff} show the average runtime and memory usage for different number of threads, over the set of multipliers shown in Tables \ref{tbl:mas-verification} and \ref{tbl:mont-verification}  (32 to 283 bits). The vertical axis on the left is CPU runtime (in seconds), and on the right is memory usage (MB). The horizontal axis represents the number of threads $T$, ranging from 1 to 30. 
The runtime is significantly improved for $T$ ranging from 5 to 15. However, there is not much speedup when $T$ is greater than 20, most likely due to the memory management synchronization overhead between the threads.
Similarly to the results for Mastrovito multipliers (Table \ref{tbl:mas-verification}), our approach is limited here by the memory usage when the size of the multiplier and the number of threads $T$ are large. In our work, $T=20$ seems to be the best choice. Obviously, $T$ varies for different platforms, depending on the number of cores and the memory. 

\begin{figure}[!htb] 
\begin{center}
\includegraphics[scale=0.40]{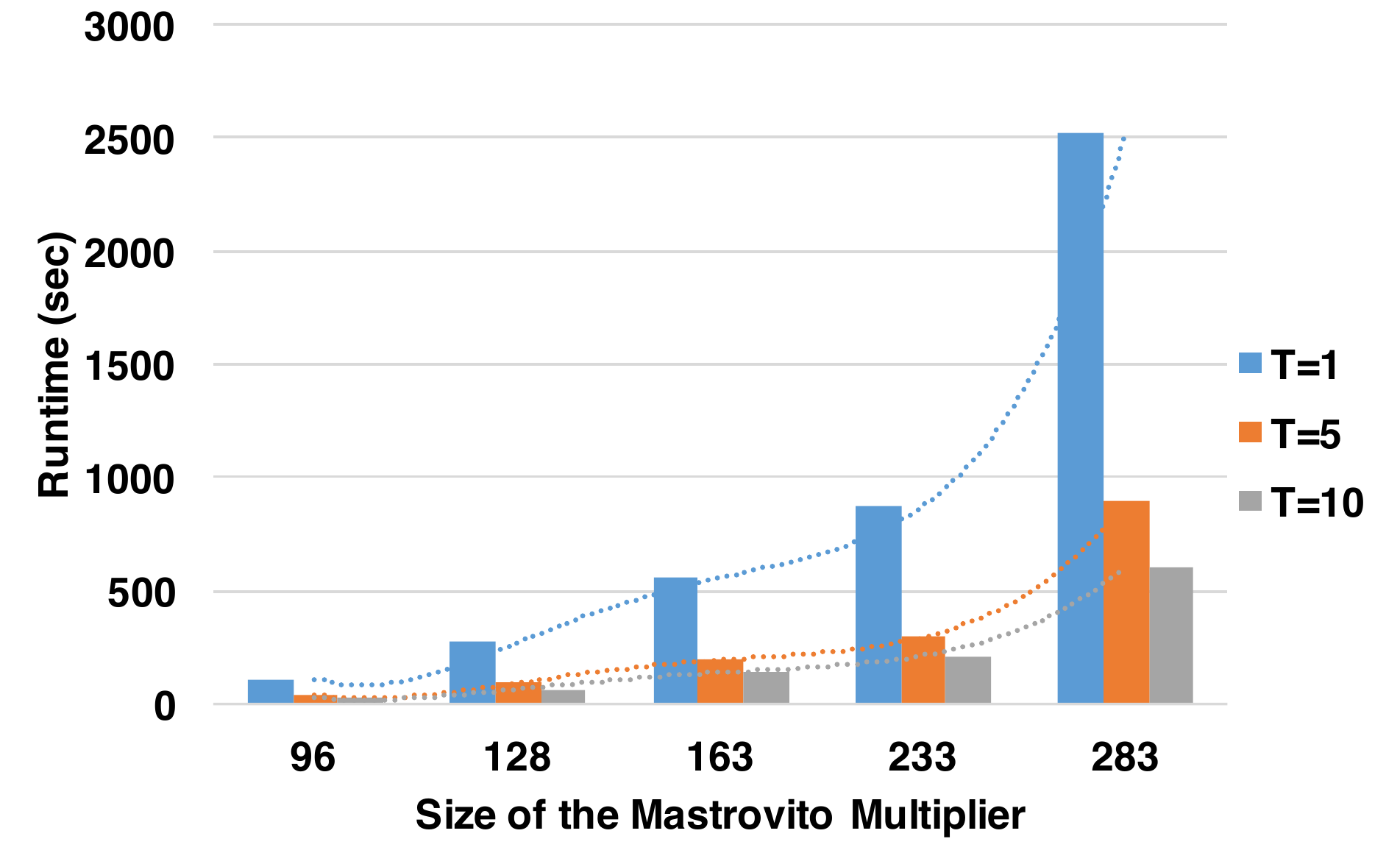}
\caption{\small{Single thread runtime analysis for Mastrovito multipliers.}}
\vspace{-4mm}
\label{fig:analysis_single}
\end{center}
\end{figure}

We also analyzed the runtime complexity of our verification algorithm for a single thread (T=1) computation; it is shown in Figure \ref{fig:analysis_single}. The y-axis shows the total runtime of rewriting the polynomial expressions, and x-axis indicates the size of the Mastrovito multiplier. 
{The result shows that the overall speedup is roughly the same for each value of T. Montgomery multipliers exhibit similar behavior, regardless of the choice of the irreducible polynomial.}

\subsubsection{\textbf{Effect of Synthesis on Verification}}

In \cite{yu:2016-tcad-verification} the authors conclude that highly bit-optimized integer arithmetic circuits are harder to verify than their original, pre-synthesized netlists. This is because the efficiency of the rewriting technique relies on the amount of cancellations between the different terms of the polynomial, and such cancellations strongly depend on the order in which signals are rewritten. A good ordering of signals is difficult to achieve in highly bit-optimized synthesized circuits. 

To see the effect of synthesis on parallel verification of GF circuits, we applied our approach to {\it post-synthesized} Galois field multipliers with operands up to 409 bits (571-bit multipliers could not be synthesized in a reasonable time). 
We synthesized \textit{Mastrovito} and \textit{Montgomery} multipliers using $ABC$ tool \cite{abc-link}. We repeatedly used the commands \textit{resyn2} and \textit{dch}\footnote{"dch" is the most efficient bit-optimization function in ABC.} until the number of AIG levels or nodes could not be reduced anymore. The synthesized multipliers were mapped using a 14nm technology library. The verification experiments shown in Table \ref{tbl:synth-verification} are performed by our tool with $T=20$ threads. Our tool was able to verify both 409-bit \textit{Mastrovito} and \textit{Montgomery} multipliers within just 13 minutes.
We observed that in our parallel approach Galois field multipliers are easier to be verified after optimization than in their original form. For example, the verification of a 283-bit Montgomery multiplier takes 15,300 seconds for $T=$20. After optimization, the runtime dropped to just 169.2 seconds, which means that such a verification is 90x faster than of the original implementation. The memory usage has also been reduced from 488 MB to 194 MB. In summary, in contrast to verification problems of integer multipliers \cite{yu:2016-tcad-verification}, the bit-level optimization actually reduces the complexity of backward rewriting process. This is because extracting the function of an output bit of a GF multiplier depends only on the logic cone of that bit and does not require logic expression from other bits to be simplified (c.f. Theorem 3). Hence, the complexity of function extraction is naturally reduced if the logic cone is minimized. 

\begin{table}[!htb]
\scriptsize
\centering
\begin{tabular}{|r|r|r|r|r|r|r|}
\hline
\multirow{2}{*}{\textit{Op size}} & \multicolumn{3}{c|}{\textit{Mastrovito}} & \multicolumn{3}{c|}{\textit{Montgomery}} \\ \cline{2-7} 
 &\# eqn & Runtime(s) & Mem &\# eqn & Runtime(s) & Mem \\ \hline
64 & 11499 & 4 & 21 MB &9471 & 15 & 38 MB\\ \hline
96 & 25632& 11 & 44 MB &20306 & 41 & 54 MB\\ \hline
128 & 45983& 29 & 77 MB &35082 & 27 & 78 MB\\ \hline
163 & 73483& 62 & 123 MB & 56408& 205 & 153 MB\\ \hline
233 & 121861& 135 & 201 MB & 110947& 141 & 199 MB\\ \hline
283 & 120877& 168 & 198 MB & 111006& 169 & 194 MB\\ \hline
409 & 385974& 776 & 635 MB &340076 & 751 & 597 MB\\ \hline
\end{tabular}
\caption{Runtime and memory usage of synthesized \textit{Mastrovito} and \textit{Montgomery} multipliers ($T$=20).}
\label{tbl:synth-verification}
\end{table}

\subsection{Reverse Engineering of GF($2^m$) Multipliers} \label{results-reverse-engin}

The reverse engineering technique presented in this paper was implemented in the framework described in Section \ref{sec:reverse-engin} in C++. It reverse engineers bit-blasted GF($2^m$) multipliers by analyzing the algebraic expressions of each element using the approach presented in Section \ref{sec:parallel_GF}. The program was tested on a number of gate-level $GF(2^{m})$ multipliers with different irreducible polynomials, including Montgomery multipliers and Mastrovito multipliers. The multiplier generator, taken from \cite{kalla:tcad13}, takes the bit-width and the irreducible polynomial as inputs and generates the multipliers in the equation format. The experimental results show that our technique can successfully reverse engineer various GF($2^m$) multipliers, regardless of the GF($2^m$) algorithm and the irreducible polynomials. We set the number of threads to 16 for all the reverse engineering evaluations in this section. This is dictated by the fact that T=16 gives most promising performance (runtime) and scalability (memory usage) metrics on our platform, based on the analysis presented in Section \ref{sec:parallel_analysis} (Figure \ref{fig:tradeoff}).


\begin{table}[!htb]
\scriptsize
\centering
\begin{tabular}{|r|r|r|r|r|r|}
\hline
\multirow{2}{*}{$m$} & \multicolumn{1}{c|}{\multirow{2}{*}{$P(x)$}} & \multicolumn{2}{c|}{\textit{Mastrovito-syn}} & \multicolumn{2}{c|}{\textit{Montgomery-syn}} \\ \cline{3-6} 
 & \multicolumn{1}{c|}{} & T(s) & Mem & T(s) & Mem \\ \hline
64 & $x^{64}$+$x^{21}$+$x^{19}$+$x^{4}$+1 & 13 & 25 MB & 5 & 20 MB \\ \hline
163 & $x^{163}$+$x^{80}$+$x^{47}$+$x^9$+1 & 69 & 508 MB & 221 & 610 MB \\ \hline
233 & $x^{233}$+$x^{74}$+1 & 152 & 1.2 GB & 154 & 2.9 GB \\ \hline
409 & $x^{409}$+$x^{87}$+1 & 825 & 6.5 GB & 855 & 10.3 GB \\ \hline
\end{tabular}
\caption{\small Results of reverse engineering synthesized and technology mapped Mastrovito and Montgomery multipliers.}
\label{tbl:syn}
\end{table}

Our program takes the netlist/equations of the GF($2^m$) implementations, and the number of threads as input. Hence, the users can adjust the parallel efforts depending on the limitation of the machines. In this work, all results are performed in 16 threads. Typical designs that require reverse engineering are those that have been bit-optimized and mapped using a standard-cell library. Hence, we apply our technique to the bit-optimized Mastrovito and Montgomery multipliers (Table \ref{tbl:syn}). For the purpose of our experiments, the multipliers are optimized and mapped using ABC \cite{abc-link}. Compared to the verification runtime of synthesized multipliers (Table \ref{tbl:synth-verification}), the CPU time spent on analyzing the extracted expressions for reverse engineering is less than 10\% of the extraction process. This is because most computations of reverse engineering approach are associated with those for extracting the algebraic expressions, as presented in Section \ref{sec:parallel_analysis}, Table \ref{tbl:synth-verification}. 

\begin{figure}[!htb] 
\begin{center}
\includegraphics[scale=0.34]{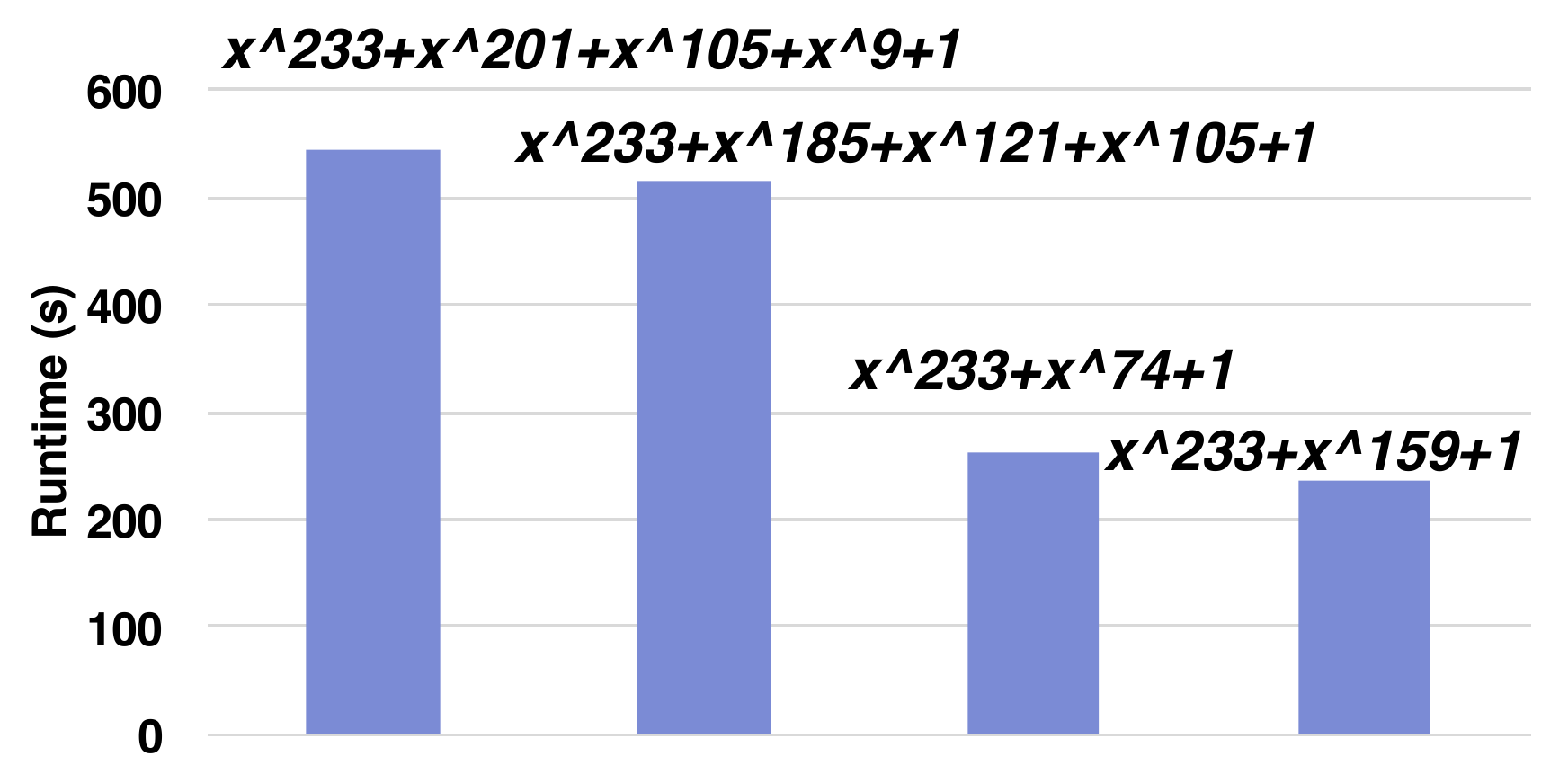}
\caption{\small Result of reverse engineering GF($2^{233}$) Mastrovito multipliers implemented with different P(x).}
\vspace{-5mm}
\label{fig:other_px}
\end{center}
\end{figure}

The reverse engineering approach has been further evaluated using four Mastrovito multipliers, each implemented with a different irreducible polynomial $P(x)$ in GF($2^{233}$). The polynomials are obtained from \cite{scott2007optimal} and optimized using ABC synthesis tool. The results are shown in Figure \ref{fig:other_px}. We can see that the multipliers implemented with trinomial $P(x)$ are much easier to be reverse engineered than those based on a pentanomial $P(x)$. This is because the multipliers implemented with pentanomial $P(x)$ contain more gates and have longer critical path, since the reduction over pentanomial requires more XOR operations. The CPU runtime for irreducible polynomial of the same class (trinomials or pentanomials) is almost the same. As discussed in Section \ref{sec:irreducible_poly}, comparison of the two trinomials shows that the efficient trinomial irreducible polynomial, $x^m$+$x^a$+1, typically satisfies $m$-$a$$>$$m/2$. 

\begin{figure}[!htb] 
\begin{center}
\includegraphics[scale=0.34]{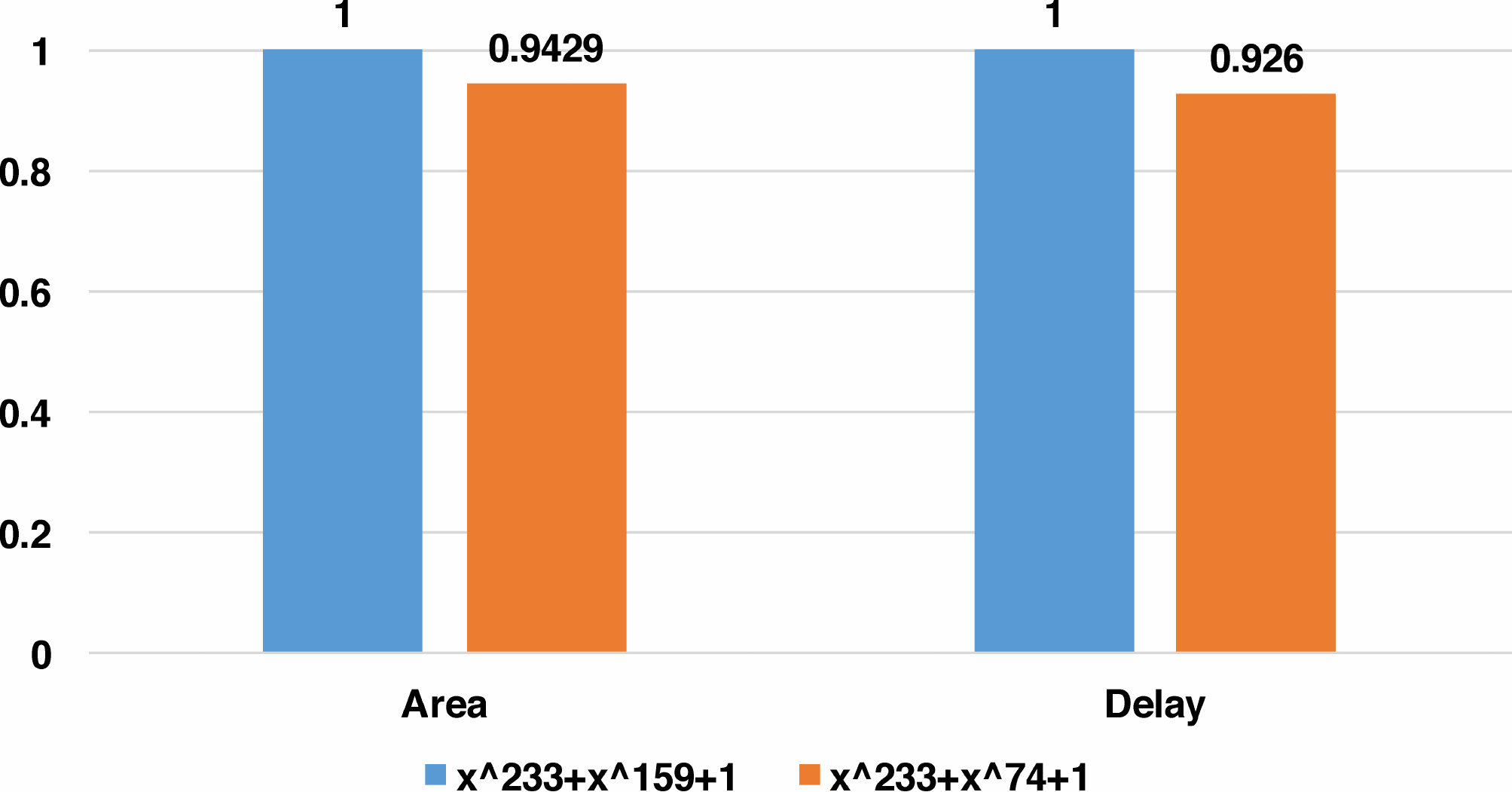}
\caption{\small Evaluation of the design cost using GF($2^233$) Mastrovito multipliers with irreducible polynomials $x^{233}$+$x^{159}$+$1$ and $x^{233}$+$x^{74}$+$1$.}
\vspace{-8mm}
\label{fig:analysis_design_cost}
\end{center}
\end{figure}

The results 
for designs synthesized with 14nm technology library are shown in Figure \ref{fig:analysis_design_cost}. It shows that the area and delay of the Mastrovito multiplier implemented with $P(x)$=$x^{233}$+$x^{74}$+$1$ are 5.7\% and 7.4\% less than for $P(x)$=$x^{233}$+$x^{159}$+$1$, respectively.

\section{Conclusion}

This paper presents a parallel approach to verification and reverse engineering of gate-level Galois Field multipliers using computer algebraic approach. It introduces a parallel rewriting method that can efficiently extract functional specification of Galois Field multipliers as polynomial expressions. We demonstrate that compared to the best known algorithms, our approach tested on $T$=30 threads provides on average 44$\times$ and 17$\times$ speedup in verification of Montgomery and Mastrovito multipliers, respectively. 
We presented a novel approach that reverse engineers the gate-level Galois Field multipliers, in which the irreducible polynomial, as well as the bit position of the inputs and outputs are unknown. We demonstrated that our approach can efficiently reverse engineer the Galois Field multipliers implemented using different irreducible polynomials. Future work will focus on formal verification of prime field arithmetic circuits and complex cryptography circuits.

\ifCLASSOPTIONcompsoc
  \section*{Acknowledgments}
\else
  \section*{Acknowledgment}
\fi

The authors would like to thank Prof. Kalla, University of Utah, for his valuable comments and the benchmarks; and Dr. Arnaud Tisserand, University Rennes 1 ENSSAT, for his valuable discussion. This work has been funded by NSF grants, CCF-1319496 and CCF-1617708.

\ifCLASSOPTIONcaptionsoff
  \newpage
\fi

\bibliographystyle{IEEEtran}
\bibliography{verification_ycunxi}
\end{document}